\documentclass[10pt]{article}

\usepackage{amsmath}
\usepackage{amssymb}

\usepackage{graphicx}

\usepackage{cite}

\usepackage{color} 

\usepackage{setspace} 
\usepackage[labelfont=bf,labelsep=period,justification=raggedright]{caption}

\makeatletter
\renewcommand{\@biblabel}[1]{\quad#1.}
\makeatother

\date{}

\begin{document}

\begin{flushleft}
{\Large
\textbf{A Computational Methodology to Screen Activities of Enzyme Variants}
}
\\
Martin R. Hediger$^{1}$, 
Luca De Vico$^{1}$, 
Allan Svendsen$^{2}$, 
Werner Besenmatter$^{2}$, 
Jan H. Jensen$^{1,\ast}$
\\
\bf{1} Department of Chemistry, University of Copenhagen, Universitetsparken 5, DK-2100 Copenhagen, Denmark
\\
\bf{2} Novozymes A/S, Krogshoejvej 36, 2880 Bagsv{\ae}rd, Denmark
\\
$\ast$ Corresponding Author, Email: jhjensen@chem.ku.dk
\end{flushleft}

\section*{Abstract}
We present a fast computational method to efficiently screen enzyme activity. In the presented method, the effect of mutations on the barrier height of an enzyme-catalysed reaction can be computed within 24 hours on roughly 10 processors. The methodology is based on the PM6 and MOZYME methods as implemented in MOPAC2009, and is tested on the first step of the amide hydrolysis reaction catalyzed by \textit{Candida Antarctica} lipase B (CalB) enzyme. The barrier heights are estimated using adiabatic mapping and shown to give barrier heights to within \mbox{3kcal/mol} of \mbox{B3LYP/6-31G(d)//RHF/3-21G} results for a small model system. Relatively strict convergence criteria (0.5kcal/(mol\AA)), long NDDO cutoff distances within the MOZYME method (15\AA) and single point evaluations using conventional PM6 are needed for reliable results. The generation of mutant structure and subsequent setup of the semiempirical calculations are automated so that the effect on barrier heights can be estimated for hundreds of mutants in a matter of weeks using high performance computing.

\section*{Author Summary}

\section*{Introduction} 
Current computational studies of enzyme activity as measured by the activation free energy generally restrict their focus to the wild type enzyme and, perhaps, one or two mutants described with a comparatively high\cite{claeyssens2006high, parks2009mechanism, hermann2009high} or moderately high\cite{noodleman2004quantum, syren2011amidases, tian2009qm} level of theory. The agreements with experimental results are often impressive and these studies can provide valuable insight into the catalytic mechanism\cite{altarsha2010coupling}. However, the computational demands of these methods makes it difficult to apply them to the actual design of new enzymatic catalysts where the activity of hundres of mutants has to be evaluated. This paper describes a computational method that makes this practically possible.

In order to make the method computationally feasible, relatively approximate treatments of the wave function, structural model, dynamics and reaction path are used. Given this and the automated setup of calculations, some inaccurate results will be unavoidable. However, the intend of the method is similar to experimental high through-put screens of enzyme activity where, for example, negative results may result from issues unrelated to the intrinsic activity of the enzyme such as imperfections in the activity assay, low expression yield, protein aggregation, etc. Just like its experimental counterpart our technique is intended to identify \textit{potentially} interesting mutants for further study.

\begin{figure}[!ht]
\begin{center}
\includegraphics[width=4in]{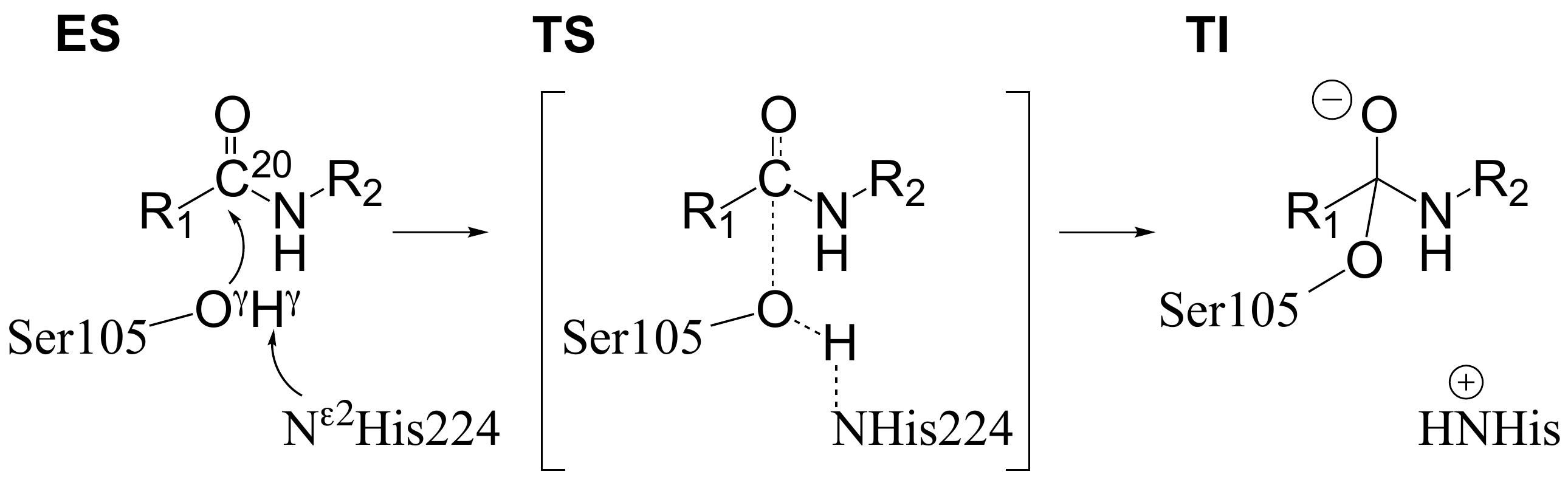}
\end{center}
\caption{{\bf Formation of tetrahedral intermediate mechanism.} Concerted nucleophilic attack of O${^\gamma}$ of S105 and abstraction of proton H$^\gamma$ by N$^{\epsilon 2}$ of H224 and development of formal negative charge on substrate oxygen. The enzyme substrate complex $\textbf{ES}$ (left) is transformed into tetrahedral intermediate, \textbf{TI}. R$_1$: \mbox{-CH$_2$CH$_3$}, R$_2$: \mbox{-CH$_2$C$_6$H$_5$}.}
\label{fig:nuc-attack}
\end{figure}

In this paper we develop and test the technique on the enzyme \textit{Candida Antarctica} lipase B (CalB) enzyme. CalB catalyses the hydrolysis of lipophilic esters and shows only very low amidase activity. While we use the method to test the effect of a few mutations on the first step in the hydrolysis of a simple amide by CalB (Fig. \ref{fig:nuc-attack}), the main point of this study is the developement of a general, efficient and robust computational method that can be used on systems similar to this.

\section*{Methods} 
In this paper we focus on estimating $k_{cat}$ rather than
$k_{cat}/K_M$ because most industrial uses of enzyme catalysts work at high
substrate concentration where $k_{cat}$ is most critical for product formation.
Therefore, like in most computational studies of enzyme catalysis, substrate
binding-affinity is not considered. The inclusion of protein dynamics is not
considered here. The most common way of estimating the effect of protein
dynamics on barrier height in QM/MM studies is to compute the barrier height
starting from different snapshots from a molecular dynamics simulation. This
way of treating protein dynamics can also be done with our method, but was not
done in this study mainly for reasons of efficiency but also because it has
not been conclusively demonstrated that this in fact increases the accuracy of
the predicted barrier. For example, Friesner and
co-workers\cite{friesner2005ab} have predicted several barrier heights within a
few kcal/mol of experiment without inclusion of such dynamic effects.
Furthermore, when estimating relatively small changes in barrier heights due to
mutations it is not clear that dynamic effects can be predicted precisely
enough from averaging over a few snapshots. However, we hope to study this
issue in future studies.

Another approximation is the use of gas phase energy
evaluations to estimate the barrier. Exploratory calculations revealed that it
is not possible to do COSMO\cite{klamt1993cosmo} calculations with
PM6\cite{stewart2007optimization} for systems as large as this using
MOPAC2009\cite{stewart1990mopac}. While it is possible to perform COSMO
calculations with MOZYME\cite{stewart1996application}, our work shows that it
is not clear that MOZYME energies are sufficiently accurate to estimate
relatively small differences in barrier height.

As will be discussed in more detail in the results section, a computational technique aimed at the study of activity in enzymes requires the molecular models to include a significant part of the enzyme. These models are in general too large to be treated with $\textit{ab initio}$ methods. The full quantum mechanical treatment of a large molecular model is however possible when using semiempirical (SE) methods in combination with linear scaling techniques. A range of semiempirical methods is therefore evaluated and discussed. In particular, the AM1\cite{dewar1985development}, PM3\cite{stewart1989optimization} and RM1\cite{rocha2006rm1} methods as implemented in the GAMESS\cite{schmidt1993general} program and the PM6\cite{stewart2007optimization} method as implemented in the MOPAC2009\cite{stewart1990mopac} program are evaluated. In the evaluation of the semiempirical methods, single point energy calculations are carried out at the B3LYP/6-31G(d) level of theory (as implemented in GAMESS). Electronic energies and enthalpy of formation, $\Delta_f$H, are not corrected for zero point energy (ZPE).

Since the semiempirical methods use a predefined (Slater type) basis set (minimal basis for AM1, PM3 and RM1, augmented by d-orbitals on main-group atoms in PM6) and core approximation\cite{dewar1977ground}, a quantum chemical geometry optimization is mainly configured by the setting of the gradient convergence criterion (GCC). When using localized molecular orbitals (LMOs) provided through the MOZYME\cite{stewart1996application} method in MOPAC, it is in addition possible to adjust the distance at which the neglect of diatomic differential overlap (NDDO) approximations\cite{pople1970approximate} are discarded and replaced by point charge interactions.

Initially, the MOZYME method generates a Lewis structure of the molecule which is used to calculate the initial density matrix for the self-consistent field (SCF) procedure. The implications of using MOZYME LMOs are further discussed below.

This work is considered only with the estimation of the barrier of the reaction of Fig. \ref{fig:nuc-attack}, whereas binding effects and solvation effects are not considered explicitly. The description of a robust and efficient technique for the estimation of said reaction barrier is the purpose of this publication.

The computer scripts for generating the molecular models are available online, the URL is 
provided in the supporting material.

\section*{Results and Discussion}\label{sec:resl} 
\subsection*{Evaluation of SE Methods}\label{sec:sub_cmdt_eval}
To assess which computational method is best suited for use in a screening approach, the first step is the evaluation of the accuracy of the various methods in predicting the geometry of the transition state (\textbf{TS}) of the reaction in Fig. \ref{fig:nuc-attack}.

\begin{figure}[!ht]
\begin{center}
\includegraphics[width=4in]{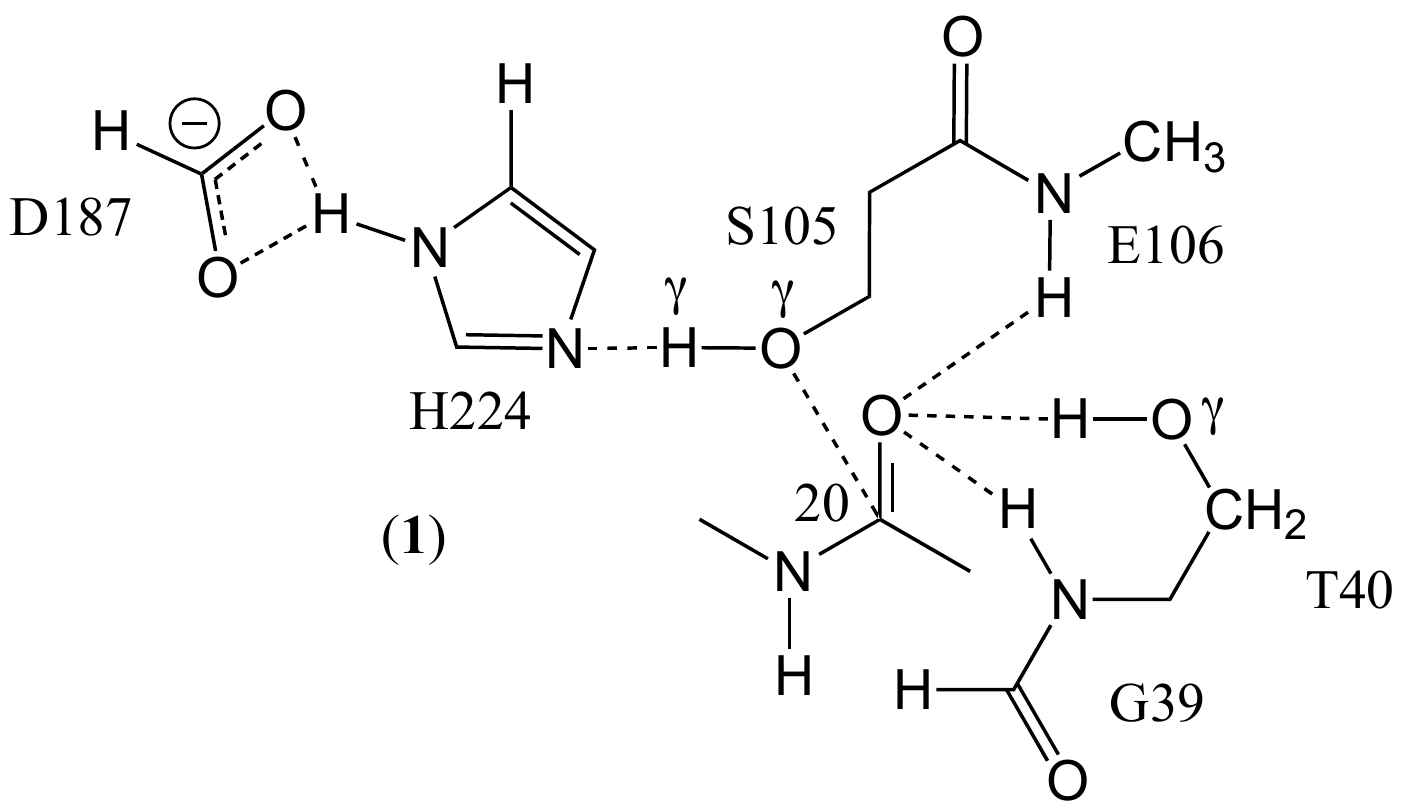}
\end{center}
\caption{{\bf Molecular system used in method evaluation.} The system carries overall charge of -1. The oxyanion hole is formed by backbone amide of G39, T40 and E106 and O$^\gamma$ of T40.}
\label{fig:chem-syst-0054}
\end{figure}

\begin{figure}[!ht]
\begin{center}
\includegraphics[width=4in]{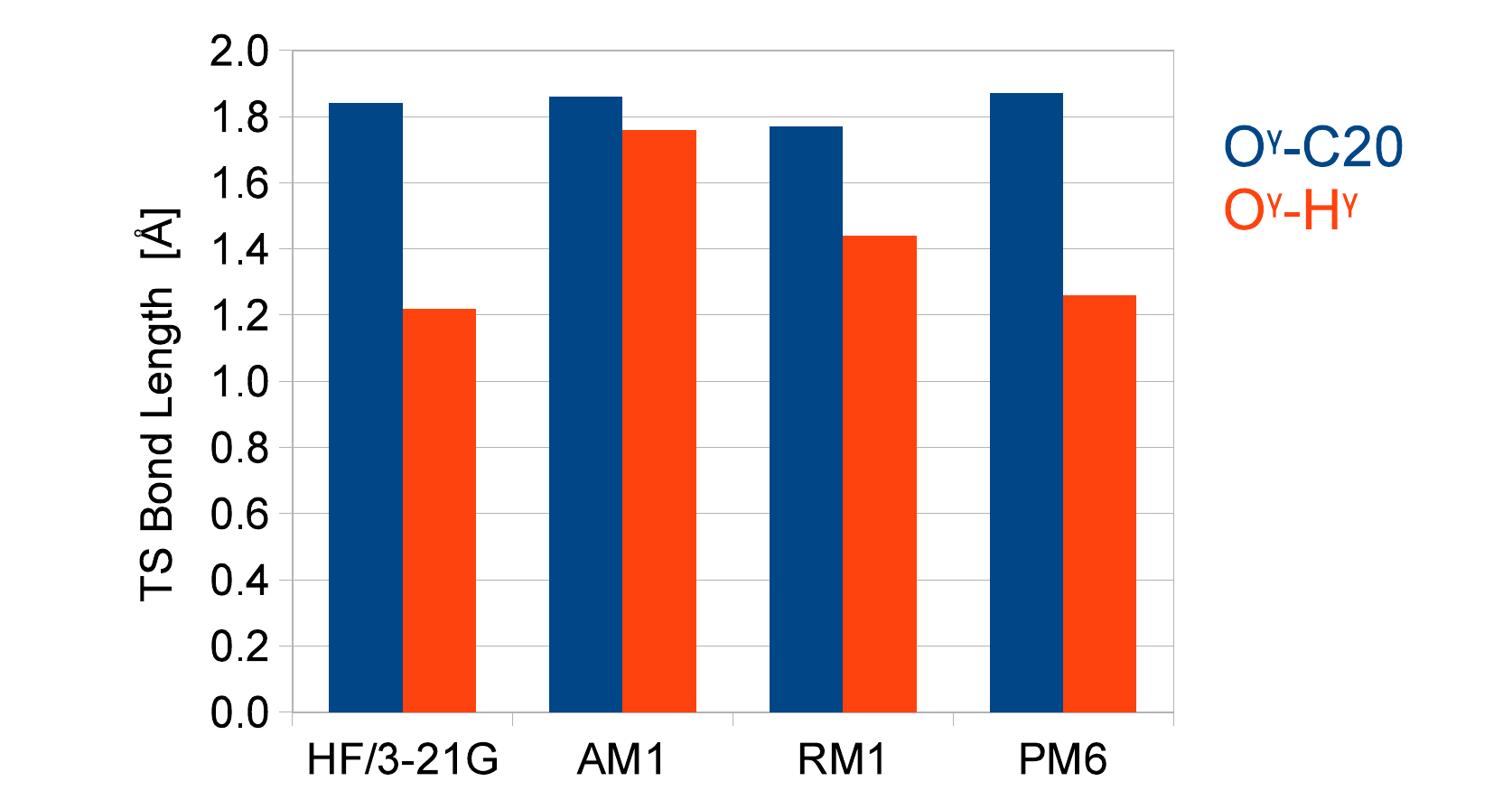}
\end{center}
\caption{{\bf TS geometries of SE methods.} Comparison of TS bond length $r$ between S105 O$^\gamma$ and H$^\gamma$ (in \AA, PM3 values not reported, see text) in (\textbf{1}). HF: r$^{O^\gamma/H^\gamma}$=1.22; PM6: r$^{O^\gamma/H^\gamma}$=1.26; AM1: r$^{O^\gamma/H^\gamma}$=1.76, i.e. H$^\gamma$ completely transfered to imidazole ring; RM1: r$^{O^\gamma/H^\gamma}$=1.44. RMSD of alignment (in \AA): HF/RM1 = 0.554; HF/AM1 = 0.538; PM6/HF = 0.224.} 
\label{fig:eval-ts}
\end{figure}

The method evaluation is done in a small model representing the active site (\textbf{AS}) of the enzyme, consisting of 54 atoms, (\textbf{1}), Fig. \ref{fig:chem-syst-0054}. The geometries of the $\textbf{TS}$ obtained from the SE methods are compared to the \textit{Hartree-Fock} (HF) geometry, Fig \ref{fig:eval-ts}.

The molecular structure of (\textbf{1}) is generated by extracting the coordinates of the atoms of the residues G39, T40, S105, E106, D187 and H224 from the crystal structure of CalB (PDB ID 1LBS\cite{uppenberg1995crystallographic}). In order to reduce computational effort, only fragments of the amino acids are included. From G39, the carbonyl group and the backbone amide is included, from T40 C$^\alpha$, C$^\beta$ and O$^\gamma$ are included. From S105, the backbone nitrogen is discarded, from E106 only the backbone nitrogen is included, the rest of the amino acid is replaced by a methyl group. D187 is represented by formic acid and from H224 only the imidazole moiety is included. All open valences are completed by the addition of hydrogens. The substrate methylacetamide (CH$_3$NHCOCH$_3$) is introduced by replacing the bound inhibitor molecule from the crystal structure.

The $\textbf{TS}$ is located by providing a suitable guess structure as a starting point followed by carrying out Newton-Raphson optimization. In the guess structure, the distance between O$^\gamma$ of S105 and C20 of the substrate is 1.80\AA\ and the distance between O$^\gamma$ and H$^\gamma$ is 1.1\AA. The $\textbf{TS}$ is located with HF and after confirmation of the $\textbf{TS}$ nature by vibrational analysis, it is used as a guess structure for the calculations with the SE methods. For every SE method, the nature of the $\textbf{TS}$ is verified by carrying out vibrational analysis. In all optimizations of \textbf{TS}, no constraints are applied. To verify that the $\textbf{TS}$ indeed connects the enzyme-substrate complex (\textbf{ES}) complex and the tetrahedral intermediate (\textbf{TI}), intrinsic reaction coordinate (IRC) calculations are carried out. The stationary end points, i.e. \textbf{ES} and \textbf{TI}, of the IRC calculation are optimized without any constraints at the same level of theory as used in the $\textbf{TS}$ search and density functional theory (DFT) single point energy calculations are performed on the optimized stationary points. In all geometry optimizations, the gradient convergence criterion is set to 0.5mHa/Bohr using GAMESS and 0.5kcal/(mol\AA) using MOPAC.

Using the distances O$^\gamma$/C20 and O$^\gamma$/H$^\gamma$ in the HF $\textbf{TS}$ and the RMSD between the HF and the SE $\textbf{TS}$ structures as a measure of comparison between different methods, it is observed that the geometry obtained from PM6 is in best agreement with the HF reference, Fig. \ref{fig:eval-ts}.

It is observed that the major difference between the methods is in the position of the H$\gamma$ proton. The distance between the nucleophilic O$^\gamma$ and C20 of the substrate is very similar in all cases.

The IRC calculations show that all methods, except PM3, are able to locate a $\textbf{TS}$ which corresponds to a concerted mechanism of nucleophilic attack and proton abstraction. The PM3 method produces a stepwise mechanism where a deprotonated serine is formed, carrying a formal negative charge. In this species, O$^\gamma$ of the serine is hydrogen bonding to the amide proton of the substrate and significant rearrangement of the molecular structure is observed (RMSD of alignment between HF and PM3: 1.66\AA).

It is observed that the energy difference for the geometries obtained by PM6 is in very close agreement to the HF reference geometry, Tab. \ref{tab:barriers}.

It is interesting to note that the energy difference based on the $\textbf{TS}$ geometry obtained from AM1 is also very close to the HF value, however, the corresponding structure is qualitatively different. Using AM1, the $\textbf{TS}$ is characterized by a deprotonated serine, whereas in HF and PM6 the proton is partially bonded between the serine and the histidine. The lower barrier from the RM1 based geometries is explained by a minor increase of the energy of the reactant relative to the \textbf{TI}.

The analysis of the $\textbf{TS}$ bond lengths and the RMSD values shows that the geometry of the $\textbf{TS}$ found with PM6 is in best agreement with the HF reference geometry. It is also noted that the PM6 method has recently been reported to provide DFT grade geometries\cite{schenker2011assessment}.

\subsection*{Molecular Enyzme Model Size}\label{sec:sub_cmdt_size}
The definition of a molecular model appropriate to use in the study of enzyme activity is subject to the following conditions. In the context of the proposed screening approach, the molecular model is required to include at least all sites which are potential targets for mutations. The upper boundary for the size of the molecular model is controlled essentially by the computational effort required for the calculation. For industrial applications, it is usually desirable to obtain results within 24 hours of wall clock time. In addition, it is assumed that the catalytic effect of a mutation located more than 10\AA\ away from the active site is negligible.

\begin{figure}[!ht]
\begin{center}
\includegraphics[width=1.0\linewidth]{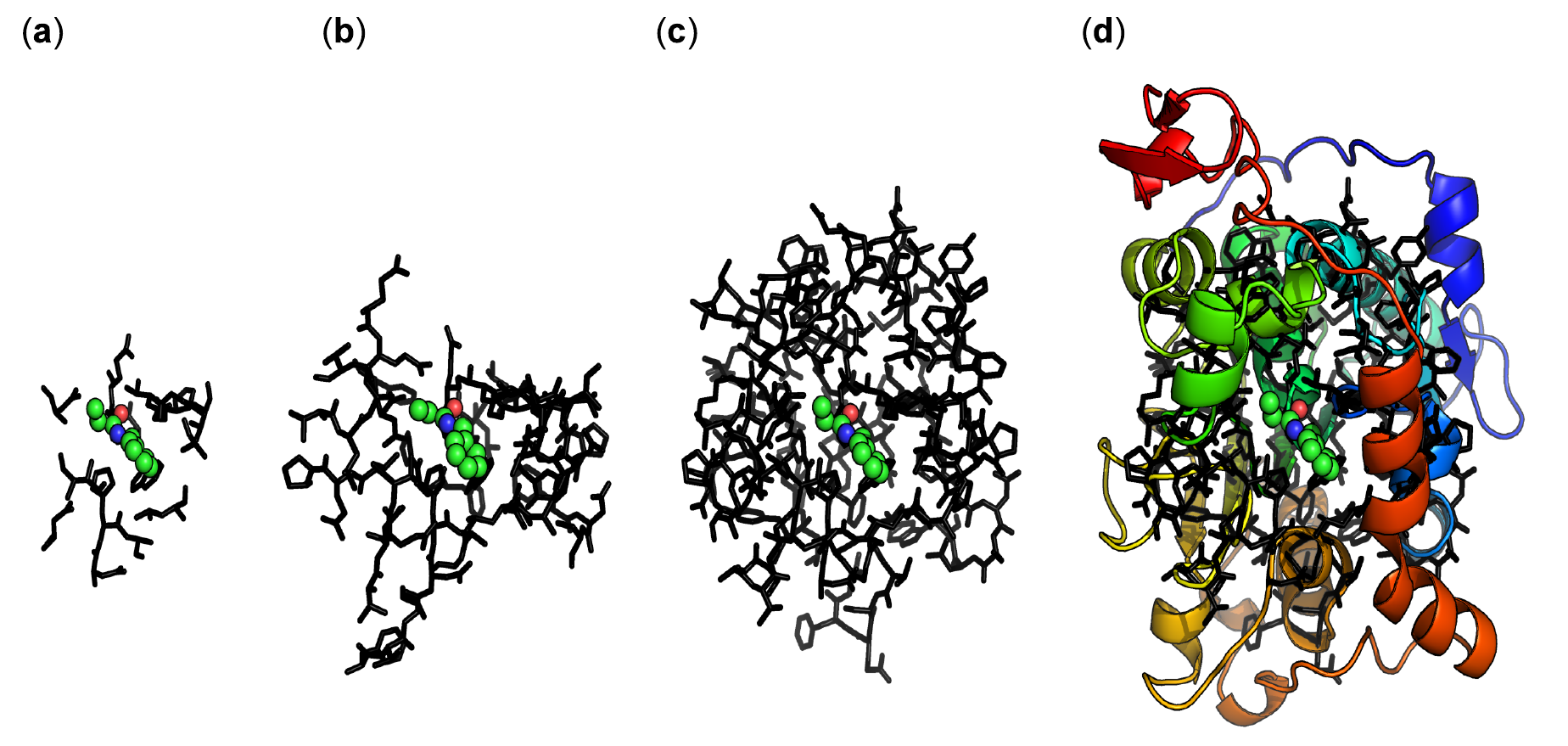}
\end{center}
\caption{{\bf Enyzme molcular model size assessment.} (\textbf{a}): 17 residues; (\textbf{b}): 55 residues; (\textbf{c}): 118 residues; (\textbf{d}): Full enzyme (316 residues) in cartoon with (\textbf{c}) overlayed in sticks. Charge on models (\textbf{a}), (\textbf{b}) and (\textbf{c}): -1, -4, -6, respectively. Protonation states of ionizable residues (at hypothetical pH of 7.4) determined with PROPKA v3.1\cite{s2011improved}, except E188 which is deprotonated.}
\label{fig:size_comparison}
\end{figure}

The molecular model and the configuration of the MOPAC program are assessed by constructing three molecular models of different sizes, Fig. \ref{fig:size_comparison}. All three models ($\textbf{a}$), ($\textbf{b}$) and ($\textbf{c}$) are based on the atomic coordinates of the crystal structure and are generated by selecting a specific set of residues (complete amino acid sequence given in the supporting material).

To afford the computational cost, the molecular model is optimized using the MOZYME LMO method and subsequent single point energy calculations are carried out using PM6 without using MOZYME. This is required since it is possible that the MOZYME energy accumulates error during geometry optimization. This observation is further discussed below.

In (\textbf{a}), only the catalytic triad, the oxyanion hole and few other residues in the active site are included. In (\textbf{b}), all residues within 8\AA\ of S105 and in (\textbf{c}) all residues within 12\AA\ of S105 are included. In case the backbone chain of the selection of residues is interrupted by only one residue, this residue is included as well. Crystal waters are also included into the molecular model. All N-termini introduced by interrupting the backbone chain are set to carry zero charge, all C-termini are modeled as -CHO groups. The benzylacetamide substrate (CH$_3$CH$_2$CONHCH$_2$C$_6$H$_5$) is introduced by molecular modeling to be in overlay with the inhibitor molecule of the crystal structure. In doing so, perfect binding is assumed. The substrate is modeled to be covalently bonded to the active site S105 and with the carbonyl carbon in tetrahedral geometry.

The effect of the MOPAC configuration is studied by optimizing the structure and computing the heat of formation, $\Delta_f$H, of the \textbf{TI}. In Tab. \ref{tab:comp_heat}, results for a set of 9 different MOPAC configurations for all three models are shown, the time requirements are further discussed below.

In (\textbf{a}), $\Delta_f$H is essentially independent of the gradient convergence criterion. This can be explained by the fact that the number of local minima is limited (compared to (\textbf{b}) and (\textbf{c})) and that gradient convergence criterion of 5kcal/(mol\AA) is sufficiently strict to lead to an optimization of all local minima. It is also observed, that the computed $\Delta_f$H does not significantly change when optimizing the structure using NDDO cutoff of 12 or 15\AA.

\begin{figure}[!ht]
\begin{center}
\includegraphics[width=1.0\linewidth]{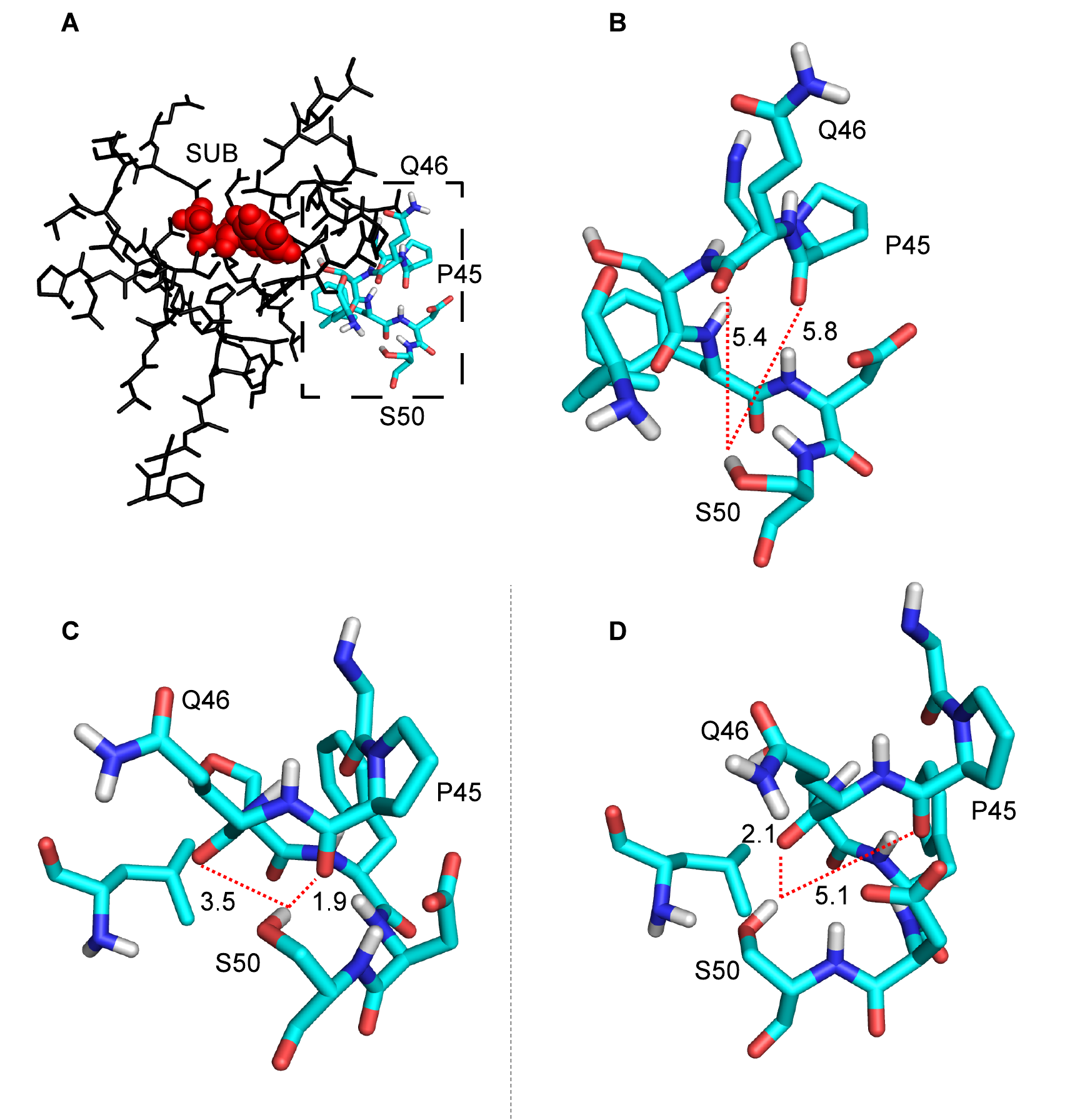}
\end{center}
\caption{{\bf Hydrogen bonding network rearrangement.} { A}. Model (\textbf{b}) overview showing the location of the residues undergoing different rearrangement in optimizations using NDDO cutoff of 12 or 15\AA, respectively. SUB: Substrate. B. Detail view of initial starting geometry. C./D. Hydrogen bonding network after optimization using NDDO cutoff of 12\AA\ or 15\AA, respectively.}
\label{fig:diff-hbon}
\end{figure}

In (\textbf{b}), significant differences in $\Delta_f$H using gradient convergence criteria of 5.0, 1.0 or 0.5 kcal/(mol\AA) are observed. It can be assumed that the strict gradient convergence criteria are required to sufficiently optimize the large number of local minima of the model, the implications of which are further discussed below. Interestingly, the optimization using gradient convergence criterion of 0.5kcal/(mol\AA) and NDDO cutoff of 12\AA\ leads to a geometry with lower $\Delta_f$H ($-4323.5$kcal/mol) compared to optimization with NDDO cutoff set to 15\AA\ ($-4318.2$kcal/(mol\AA)). The observed reason for this is that although using identical starting geometries, different NDDO cutoff settings can result in different final hydrogen bonding networks which are eventually lower in energy. This observation is made with the residue S50, which is located on the surface of model (\textbf{b}), Fig. \ref{fig:diff-hbon}a. Initially, O$^\gamma$ of S50 is roughly equally distant of the backbone carbonyl groups of P45 and Q46, Fig. \ref{fig:diff-hbon}b. After optimization, new hydrogen bonds are formed differently when optimizing with NDDO cutoff of 12 or 15\AA, respectively, Figs. \ref{fig:diff-hbon}c, d.

The rearrangement of surface residues has to be considered an inheritant artifact of the method, however it is interesting to note that different NDDO cutoff can lead to different rearrangement of the hydrogen bonding network.

It is further observed (Tab. \ref{tab:comp_heat}) that the required time to optimize the system (\textbf{b}) using strict gradient convergence criterion and NDDO cutoff is within the time frame offered in industrial environments.

Model (\textbf{c}) consists of around half of all residues of the full enzyme leading to a large number of local minima on a flat potential energy surface. A strict gradient convergence of 15kcal/(mol\AA) combined with a high NDDO cutoff distance of 15\AA\ is required to completely optimize all parts of the model. In a model of this size, $\Delta_f$H is considerably reduced both with gradient convergence and NDDO cutoff distance. Model (\textbf{c}) possibly provides the most detailed description of the active site, however, the computational time required to optimize the structure makes it prohibitive to use in a screening approach. 

\begin{figure*}[!ht]
\begin{center}
\begin{minipage}{0.32\linewidth}
\mbox{GCC: 5.0kcal/(mol\AA)}
\includegraphics[width=0.99\linewidth]{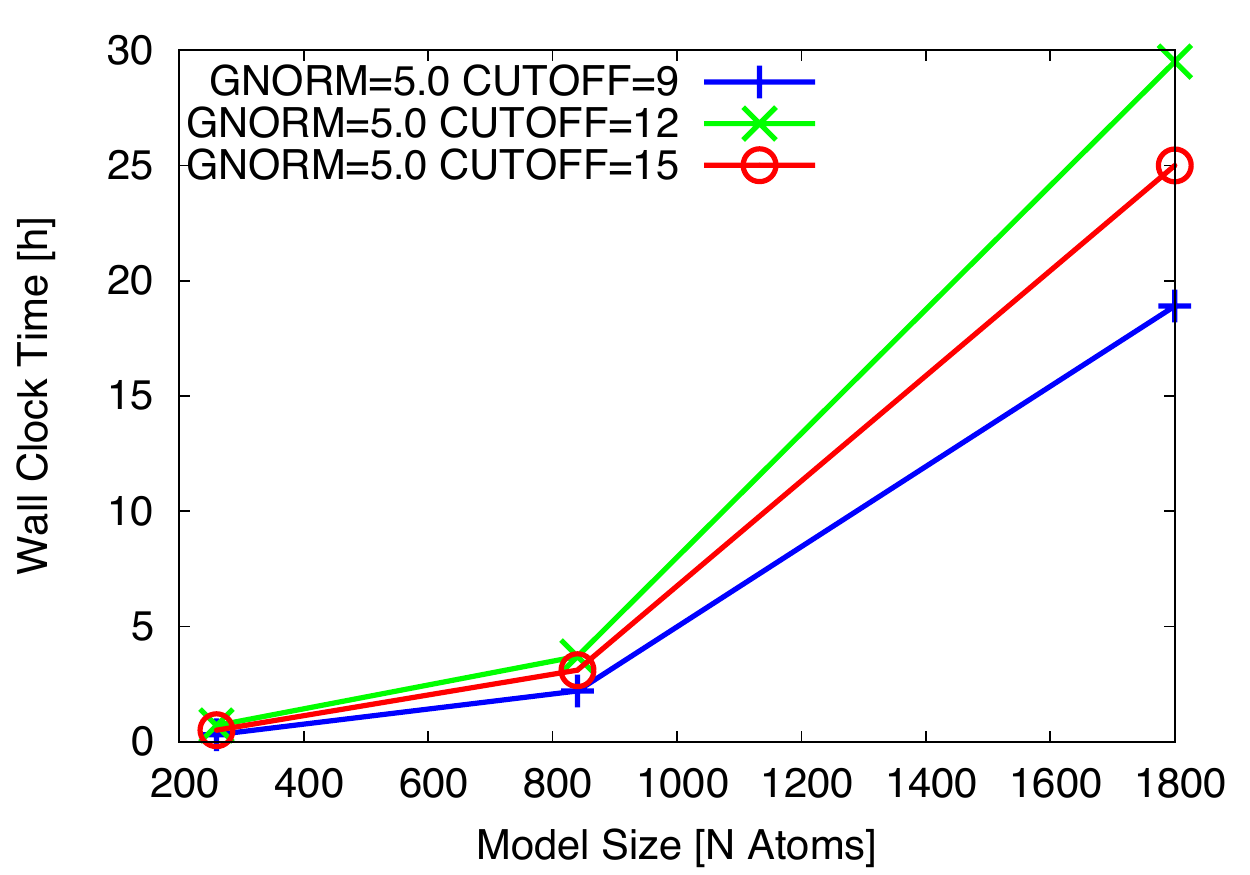}
\end{minipage}
\begin{minipage}{0.32\linewidth}
\mbox{GCC: 1.0kcal/(mol\AA)}
\includegraphics[width=0.99\linewidth]{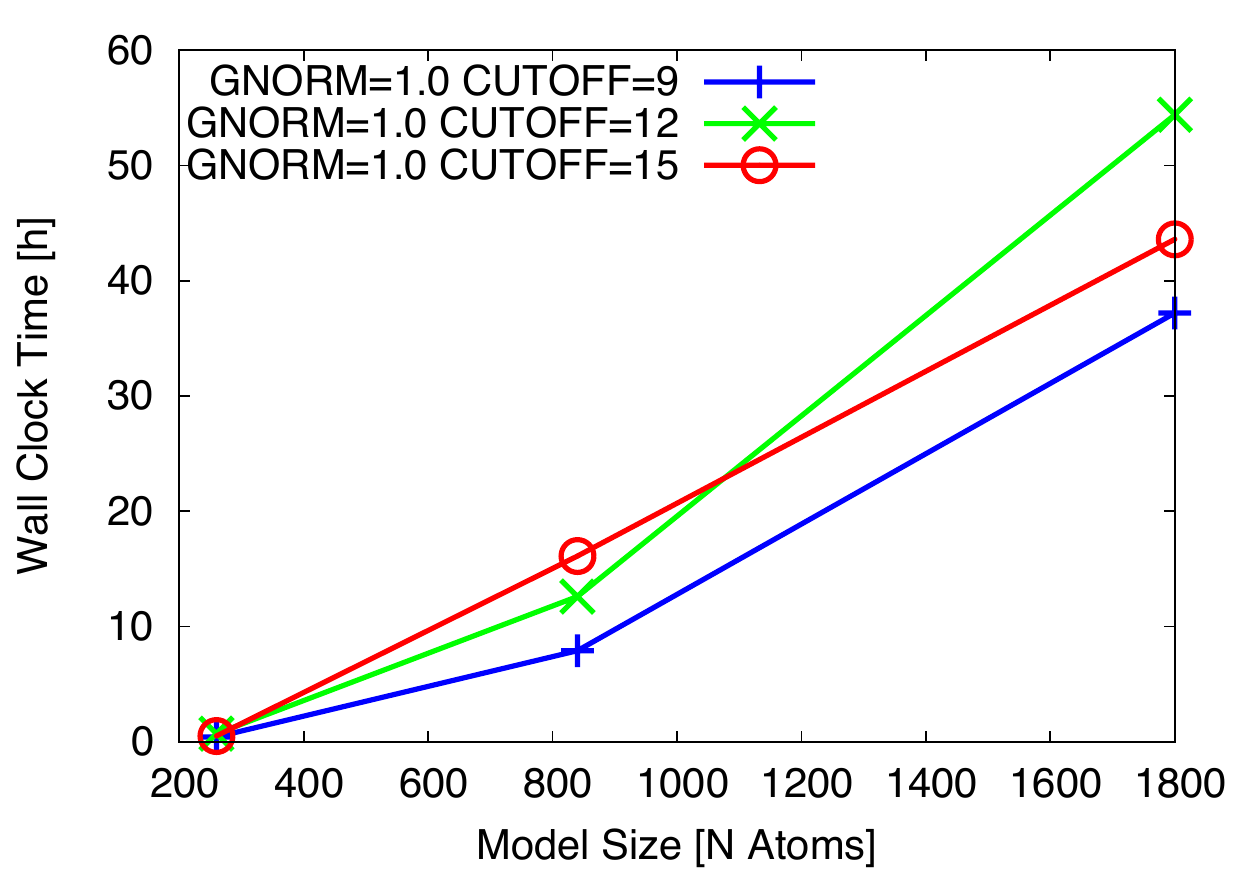}
\end{minipage}
\begin{minipage}{0.32\linewidth}
\mbox{GCC: 0.5kcal/(mol\AA)}
\includegraphics[width=0.99\linewidth]{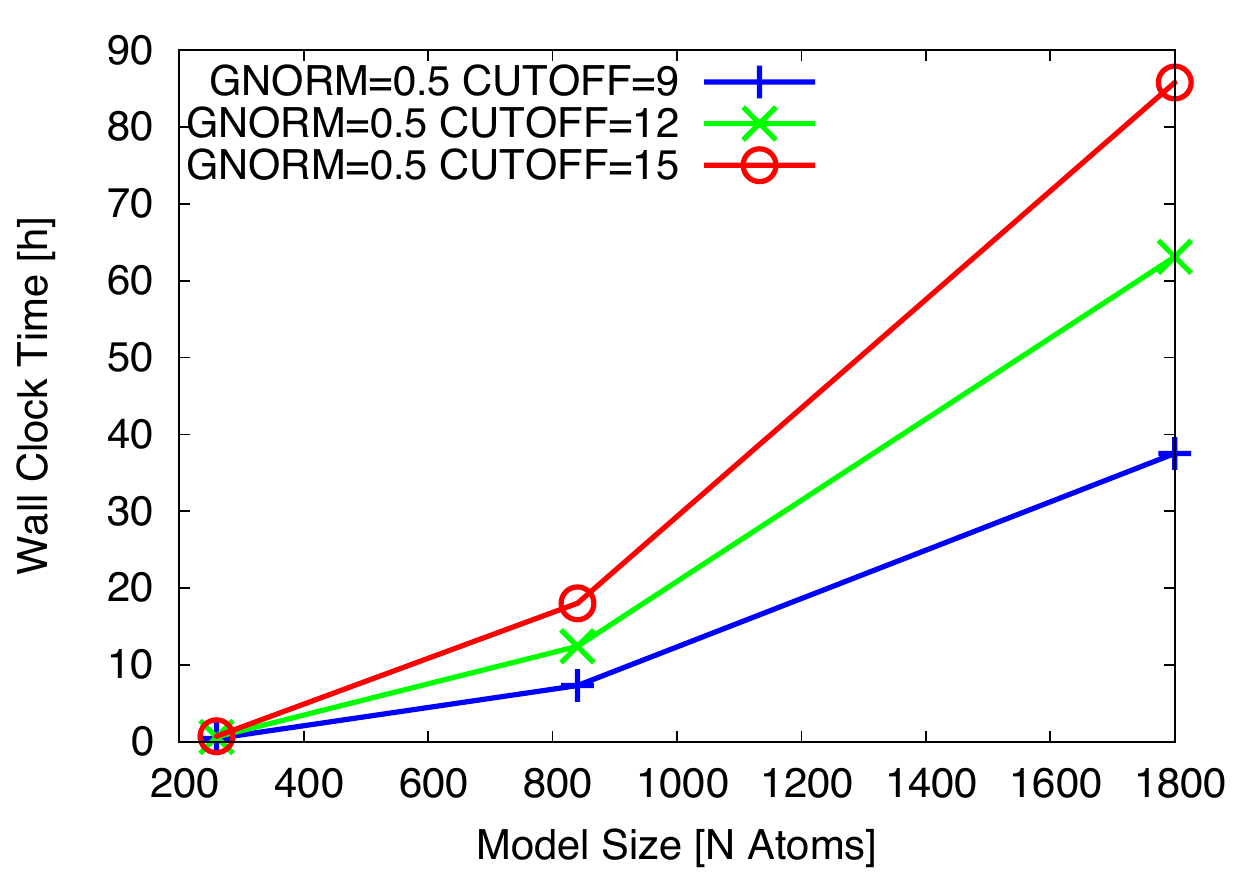}
\end{minipage}
\end{center}
\caption{{\bf Time comparison for MOPAC configuration.} { W}all clock time requirements for optimization of tetrahedral intermediate using different GCC and model sizes.}
\label{fig:time-mod-size}
\end{figure*}

The required computational wall clock time for optimization of models (\textbf{a}), (\textbf{b}) and (\textbf{c}) in dependence of gradient convergence criterion is summarized in Fig. \ref{fig:time-mod-size}.

It is observed that the required wall clock time for complete optimization of the molecular model increases non-linearly with model size. Only when using gradient convergence criterion of 1.0kcal/(mol\AA) is linear scaling of wall clock time with model size observed for NDDO cutoff of 15\AA. Using strict gradient convergence criterion of 0.5kcal/(mol\AA), linear scaling of wall clock time is approached only for NDDO cutoff distance of 9\AA.  

From considering the observed time requirements, it is concluded that an intermediately sized model like (\textbf{b}) is adequately suited for the proposed screening approach.

\subsection*{Wild Type Reaction Barrier Estimation}\label{sec:sub_cmdt_conv}
In establishing an enzyme activity screening techinque, it was tested if an approach similar to the one discussed above can be used to study activity in ($\textbf{b}$). Using the \verb+GEO_REF+ keyword\cite{stewart2009application}, the MOPAC program offers an optimization routine where two structures on either side of a reaction barrier are provided to the program. The one higher in energy is used as a reference structure for the one lower in energy. An adjustable penalty potential (based on the geometrical difference between the two structures) is then applied in the optimization of the low energy structure, which will be forced to move towards the transition state on the potential energy surface (PES). After a few cycles of optimization (using the penalty potential), in principle a guess of the transition state is obtained which can be refined using a transition state search routine. However, despite extensive testing of different magnitudes of the penalty potential, it was frequently observed that the optimization is unsuccessful in generating a valid estimate of the transition state for the reaction under consideration. Instead, the structure under the penalty potential remains on one side of the barrier or completely passes the barrier. The exact location of the transition state in large systems by this method is thus not routinely feasible and the approach is not applicable in an industrial context where semi-quantitative estimates of the overall activity are requested within one day of CPU time. This limitation becomes even more apparent when a large library of mutants is to be studied.

Based on these experiences and the results from above, it is therefore required to estimate a transition state, as described below. In the following, the notation "M(C15, G0.5)" means that a geometry optimization is carried out with the NDDO cutoff set to 15\AA\ and the gradient convergence criterion is set to 0.5kcal/(mol\AA) using the MOZYME LMO method. In this section all calculations are referring to the wild type (WT) structure.

In the procedure, first the molecular model of the \textbf{TI} is generated as described above. The \textbf{TI} model is then optimized with M(C15, G0.5). The optimized \textbf{TI} is then used as a template for the structure of the \textbf{ES} complex. To generate a model for the \textbf{ES} structure, the covalently bound substrate is replaced by the non-bonded, planar substrate and H$^\gamma$ of S105 is transferred back onto O$^\gamma$ of S105 using molecular modeling. The \textbf{ES} structure is then optimized with M(C15, G0.5). These two optimized reaction end point structures are used in the linear interpolation scheme. To assess which distance between substrate C20 and O$^\gamma$ of S105 is appropriate in the starting geometry of the \textbf{ES} complex, a number of different starting geometries were generated and optimized using M(C15, G0.5). The distance betweeen C20 and O$^\gamma$ in these starting geometries was varied in the range from 2.8 to 4.1\AA. The average distance of the optimized geometries is observed to be 3.55\AA\ and based on this, the distance between C20 and O$^\gamma$ in the starting geometry of the \textbf{ES} complex was set to 3.5\AA. No significant differences in energy were observed for the optimized geometries of the different \textbf{ES} complexes.

\begin{figure}[!ht]
\begin{center}
\includegraphics[width=4in]{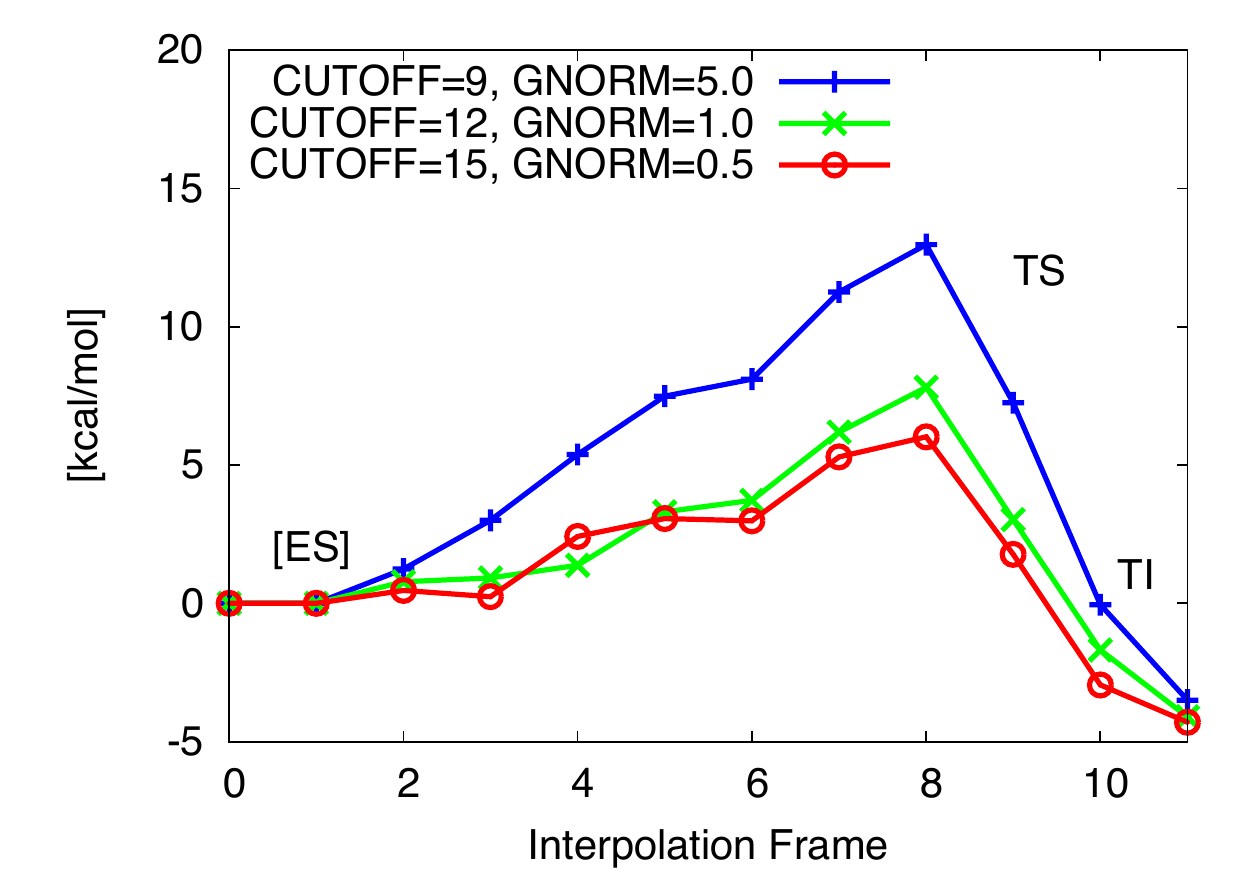}
\end{center}
\caption{{\bf WT reaction barrier estimation.} { C}onvergence of estimated reaction barrier in WT. Estimated barriers are observed to converge to a lower boundary with strict GCC and higher NDDO cutoff. All constraints discarded in first and last interpolation frame. Estimated barriers are (in \mbox{kcal/mol}) PM6//M(C9, G5.0): 13.0, PM6//M(C12, G1.0): 7.8, PM6//M(C15, G0.5): 6.0.}
\label{fig:barriers}
\end{figure}

The linear interpolation is carried out by dividing the geometrical distance between all atom pairs, $q_{i}^{\textbf{TI}} - q_{i}^{\textbf{ES}}$, where $q_i$ is any of the cartesian coordinates of the atom $i$, by 10 and adding this difference incrementaly to $q_i^{\textbf{ES}}$. Every interpolation frame generated by this procedure is then optimized with M(C15, G0.5) where in each frame, the distance between O$^\gamma$ and C20 of the substrate is kept fixed during the optimization. The separation O$^\gamma$/C20 is considered as defining the reaction coordinate and is fixed to a given value in a specific interpolation frame. The distance between C20 in the \textbf{ES} complex and C20 in the \textbf{TI} is observed to be 2.2\AA. The division of this distance by 10 interpolation steps leads to translation of C20 by 0.22\AA\ towards O$^\gamma$ of S105 in each interpolation frame. To test for convergence with MOPAC configuration, every interpolation frame is also optimized with M(C12, G1.0) and M(C09, G5.0), where the same atom pair is kept fixed during the optimization. The structure corresponding to the highest point on the obtained energy profile estimate is considered as the approximation to the \textbf{TS}. This estimate is further analysed below. The estimated barriers for three MOPAC configurations are shown in Fig. \ref{fig:barriers}.

The estimated barrier of 6.0kcal/mol (using M(C15, G0.5)) is compared to a free energy of activation of 17.8kcal/mol for the formation of tetrahedral intermediate in a high level QM/MM study\cite{ishida2003theoretical} of trypsin and 15-20kcal/mol in experimental studies\cite{fersht1999structure}. The observed difference is possibly explained by the way the \textbf{ES} complex is modeled. In our presented approach, the molecular model of the \textbf{ES} complex is based on the optimized model of the \textbf{TI}. By placing the non-covalently bonded substrate into the active site of the \textbf{TI}, a perturbation of this structure is introduced. However, the overall geometrical configuration of the active site is still very likely to the \textbf{TI} state (which itself is based on the crystal structure of the enzyme with covalently bound tetrahedral inhibitor) and therefore the optimization of the model can not completely leave the local minimum of the \textbf{TI} and arrive at the \textbf{ES} state with lower energy.

Given the very similar structure found for the enzyme-substrate
complex for virtually all mutants, the effect of using a higher energy
conformation on the barrier height will likely cancel.  As a result it will
have a relatively small effect on the relative barrier heights, which is the
key parameter in this study. However, this is another approximation invoked
to keep the method efficient.

It has to be noted that since the starting geometry for the M(C9, G5.0) and M(C12, G1.0) calculations is the optimized geometry from the M(C15, G0.5) calculation (of the stationary points), the optimized hydrogen bonding network is not expected to restructure. This is the reason why the \textbf{TI} obtained from optimizing with M(C12, G1.0) does not have the same relative energy as in Tab. \ref{tab:comp_heat}, where the structure obtained from M(C12, G1.0) is lower in energy than the one obtained from M(C15, G0.5).

The estimated barriers using M(C12, G1.0) and M(C15, G0.5) are characterized by the same shape, while the estimated barrier using M(C9, G5.0) is significantly different. The apparent difference when going from less strict to strict gradient convergence is possibly explained by the fact that the PES of the system contains a huge amount of local minima. Using strict gradient convergence, it is ensured that also those parts of the gradient corresponding to shallow local minima are minimized. This in turn is apparently responsible for quite significant lowering of overall energy of the system.

From the above, it can be concluded that using a NDDO cutoff of at least 12\AA\ and a gradient convergence criterion of at least 1.0kcal/(mol\AA) is required for converged estimation of the reaction barrier.

\subsection*{Transition State Verification}\label{sec:var_cmdt_tsvr}
\begin{figure}[!ht]
\begin{center}
\begin{minipage}{0.48\linewidth}
\includegraphics[width=0.99\linewidth]{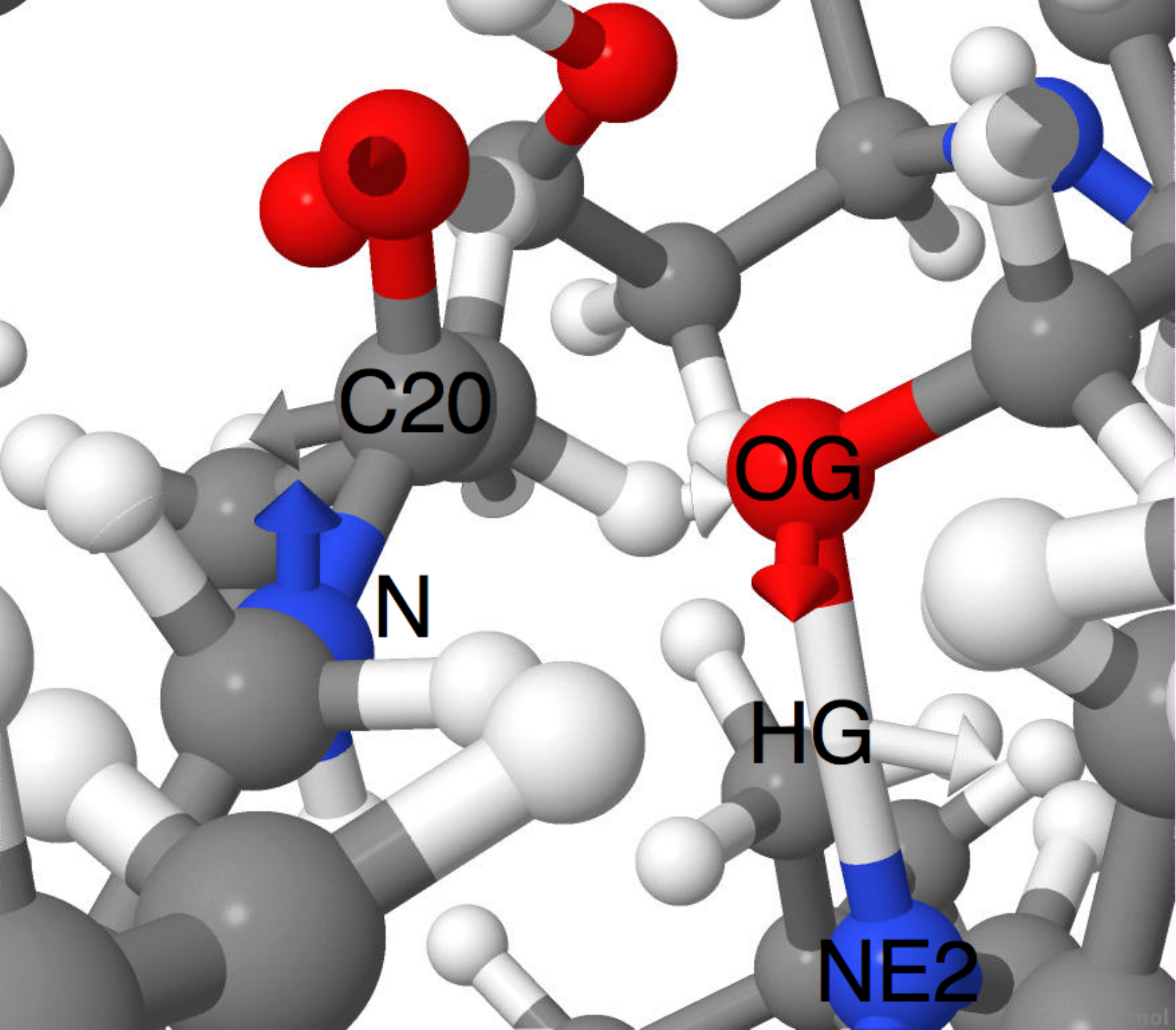}
\end{minipage}
\hfill
\begin{minipage}{0.48\linewidth}
\includegraphics[width=0.99\linewidth]{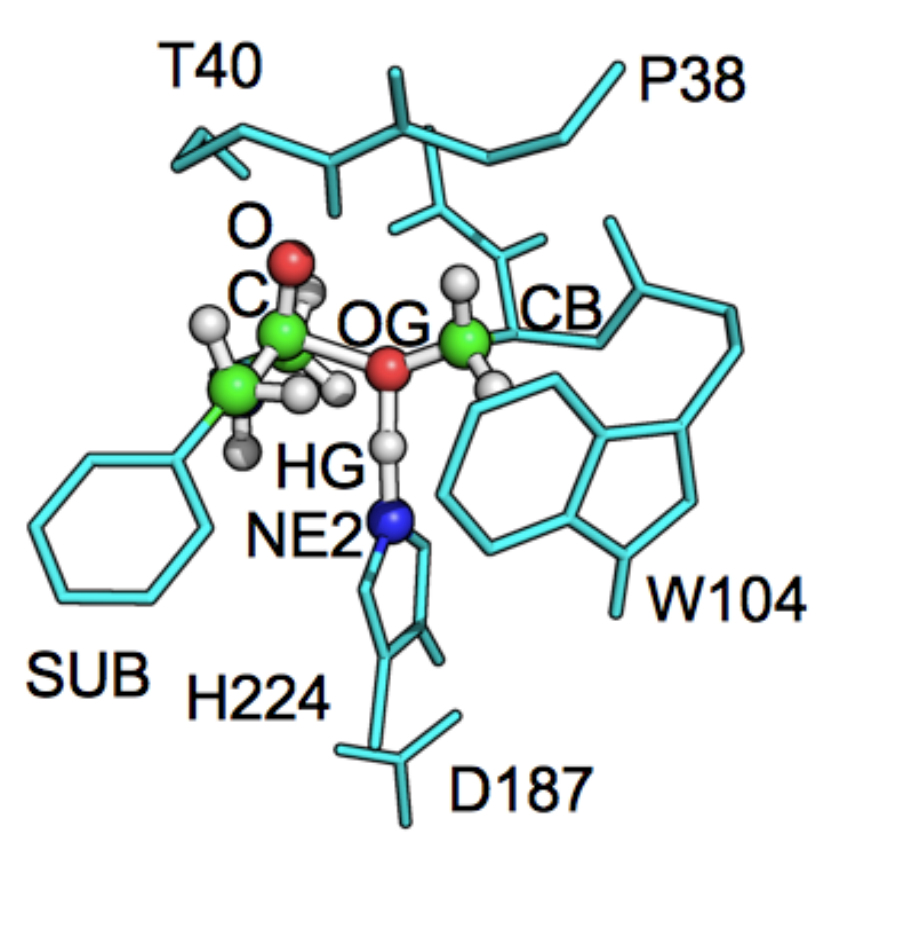}
\end{minipage}
\end{center}
\caption{{\bf Illustration of PHVA in WT \textbf{TS}.} { A}. Normal mode vibration on C20 of substrate towards S105 O$^\gamma$ (OG), H$^\gamma$ (HG) normal mode vector is towards H224 (NE2). B. Atoms included in PHVA shown as spheres (16 atoms in total). PHVA required 24.3h of wall clock time.}
\label{fig:intr-phva}
\end{figure}

The optimized interpolation frame corresponding to the highest point on the energy profile (Frame 8 in Fig. \ref{fig:barriers} of the M(C15, G0.5) calculation) is subjected to partial Hessian vibrational analysis\cite{li2002partial} (PHVA) using PM6 (without MOZYME, this function is provided by the \verb+FORCETS+ keyword in MOPAC). One imaginary frequency is found (91.9$i$cm$^{-1}$). The normal mode vibration is sketched in Fig. \ref{fig:intr-phva}. An animation of the vibration is available in the supporting material.

It has to be noted that the distance O$^\gamma$/C20 is constrained in the interpolation and results from the (arbitrary) division of the reaction coordinate into ten interpolation frames. Nevertheless, in the interpolation frame 8, the distances of S105 O$^\gamma$/C20 and O$^\gamma$/H$^\gamma$ are 1.88\AA\ (fixed) and 1.27\AA\ (optimized), respectively, which is in very close agreement to the transition state distances found in model (\textbf{1}) using PM6 (Fig. \ref{fig:eval-ts}). It can be concluded that the highest point on the reaction barrier estimate occurs at a geometry which is quite similar to the completely optimized \textbf{TS} structure of model (\textbf{1}).

Carrying out partial Hessian vibrational analysis using MOZYME LMOs returns only positive frequencies. Also it is observed, that all frequencies are positive after carrying out a partial transition state search for the atoms of the PHVA in the optimized interpolation frame 8.

\subsection*{Comparison of PM6 and MOZYME Energies}\label{sec:sub_cmdt_comp}
\begin{figure}[!ht]
\begin{center}
\includegraphics[width=4in]{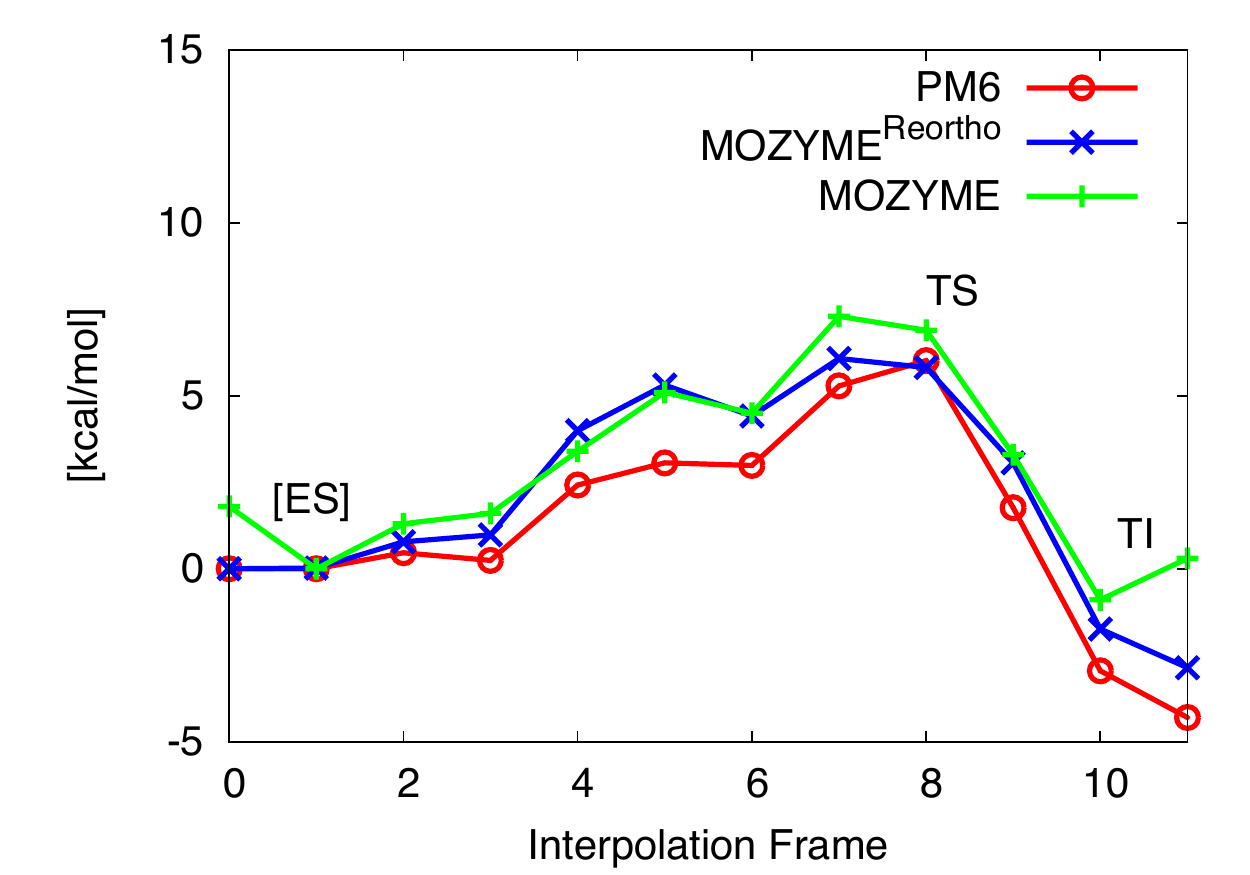}
\end{center}
\caption{{\bf Comparison of PM6 and MOZYME $\Delta_f$H.} { C}omparison of $\Delta_f$H using the final MOZYME LMOs, MOZYME LMOs after reorthogonalization and delocalized orbitals with PM6. Increases for $\Delta_f$H in frames 0 and 11 of the MOZYME curve are due to loss of orthogonality of the LMOs. Calculations involving MOZMYE are done using M(C15, G0.5). Estimated barriers are (in kcal/mol) MOZYME: 7.3, MOZYME$^{Reortho}$: 6.1, PM6: 6.0.}
\label{fig:comp_moz_ort_pm6}
\end{figure}

In the MOZYME method, in geometry optimization step $\tau_i$, where $i>0$, the LMOs from the step $\tau_{i-1}$ are used as the starting LMOs in the SCF procedure. The error originating from the truncation of the LMOs in step $\tau_{i-1}$ is therefore also present in the SCF cycle of the $\tau_i$ step. This can lead to different MOZYME and PM6 energies and differences in the estimated reaction barriers. In principle, this effect is avoided if the energy of the final geometry is evaluated using the \verb+1SCF+ keyword to form a reorthogonalized set of LMOs, see Fig \ref{fig:comp_moz_ort_pm6}.

As shown, the loss of orthogonality increases with the number of SCF cycles required in the geometry optimization. This is apparent in frames 0 and 11 of Fig. \ref{fig:comp_moz_ort_pm6}. The number of complete SCF cycles in these frames are 494 and 1896, respectively, compared to 25 (frame 1) and 437 (frame 10). Further comparisons between $\Delta_f$H values obtained using different NDDO cutoff distances compared to PM6 are given in the supporting material.

\begin{figure}[!ht]
\begin{center}
\includegraphics[width=4in]{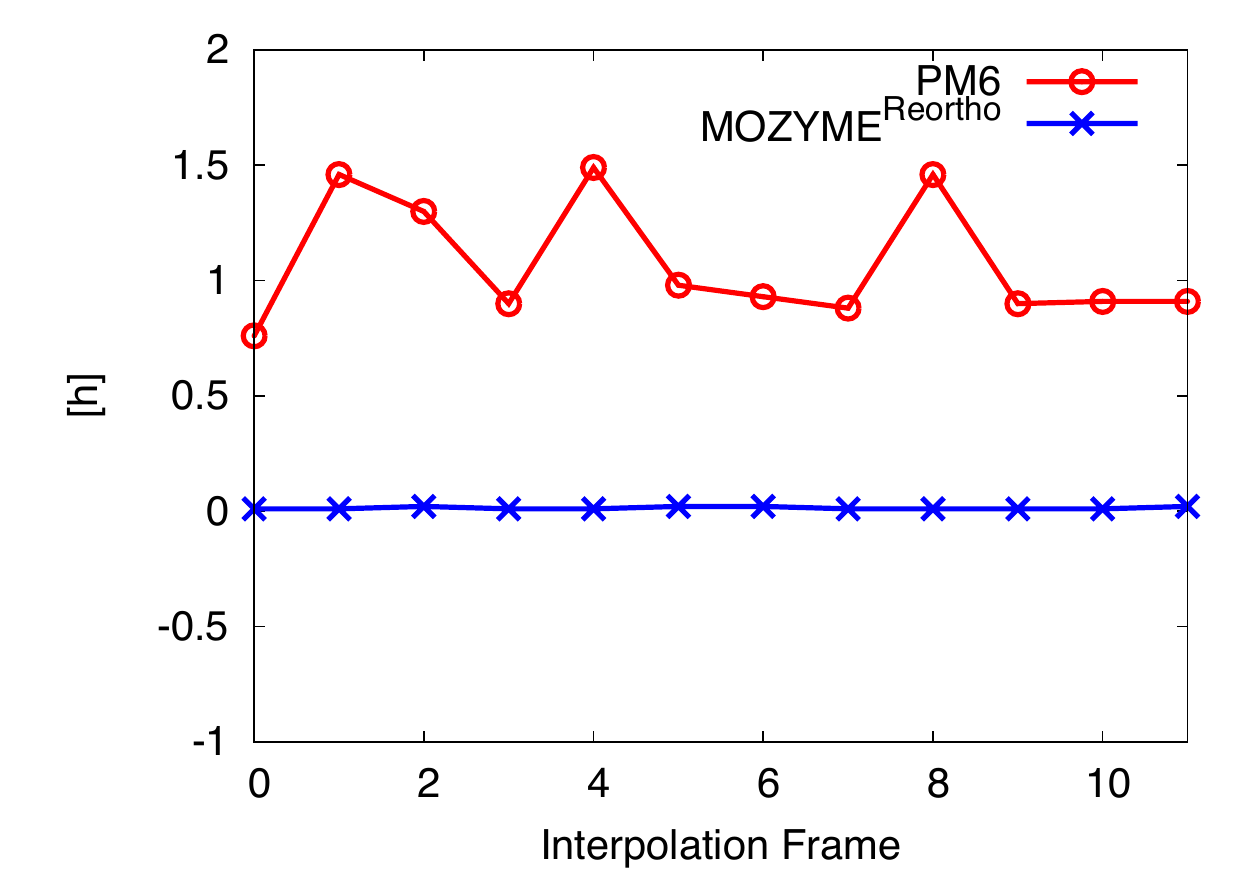}
\end{center}
\caption{{\bf Comparison of CPU time required for PM6 and MOZYME SPE calculation.} Average CPU time (in h): 1.07 (PM6), 0.01 (MOZYME).}
\label{fig:comp_cpu_moz_ort_pm6}
\end{figure}

The required computational time to calculate single point energies using MOZYME is significantly different compared to using non-localized MOs, see Fig. \ref{fig:comp_cpu_moz_ort_pm6}.

\subsection*{Variant Model Preparation and Single Mutation Screening}
In the optimized model of the stationary points of the wild type, the molecular model of the variant \textit{v} is generated by mutating the respective position in the backbone using the PyMOL\cite{PyMOLu} \textit{Mutagenesis Wizard} function. The two molecular models (\textbf{ES}$^v$ and \textbf{TI}$^v$) are then used in a similar linear interpolation scheme as described for the wild type above. To illustrate the approach, the (single) mutations G39A, T103G and W104F are studied. Of the three discussed variants, G39A and W104F are located in the active site, C$^\alpha$ of T103G is located 8.7\AA\ away from O$^\gamma$ of S105.

\begin{figure}[!ht]
\begin{center}
\includegraphics[width=4in]{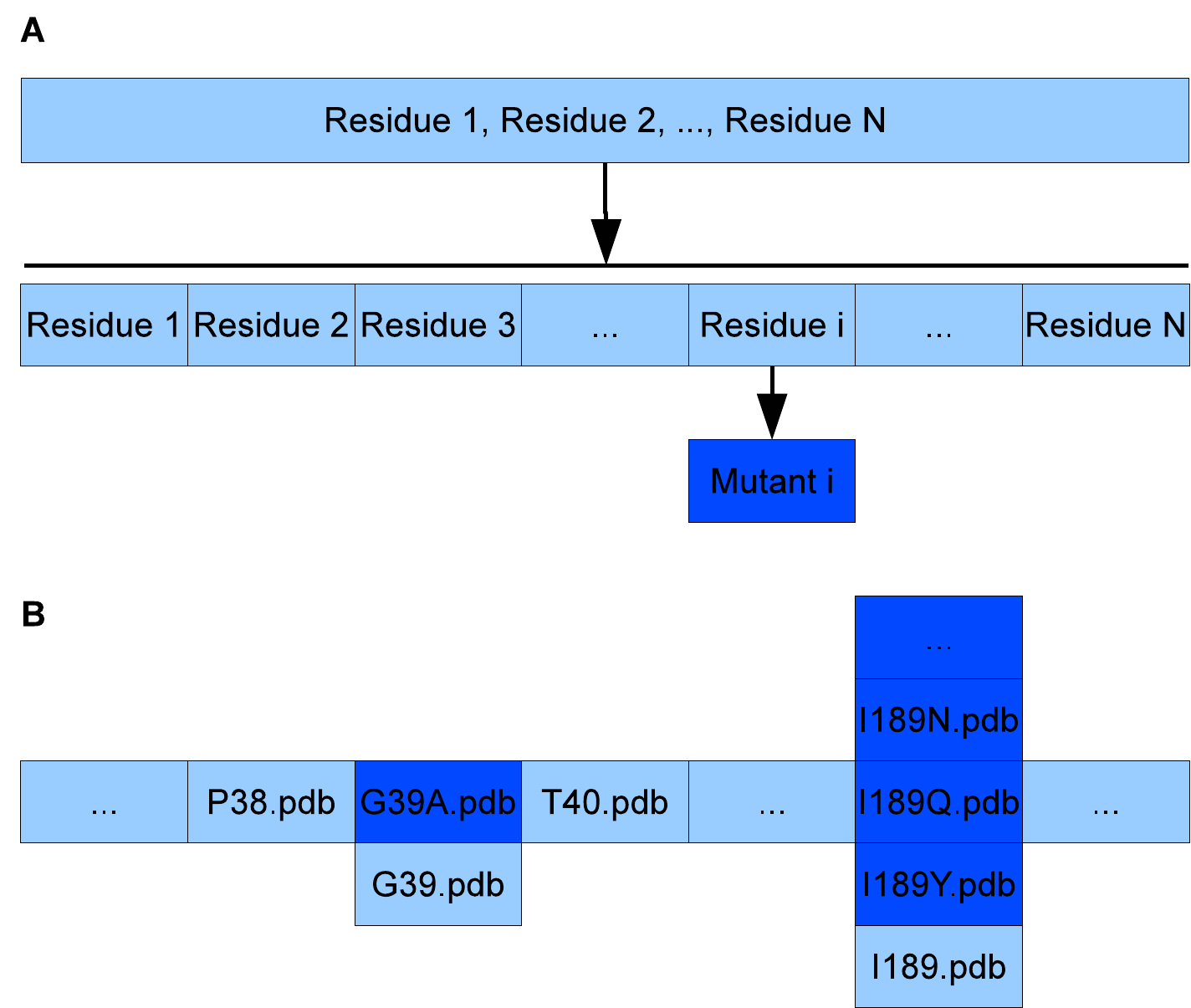}
\end{center}
\caption{{\bf Mutant file assembly.} { A}. Preparation of fragment PDB files of WT and mutant. B. Sketch of variant structure file assembly from fragment PDB files using CalB. Light blue boxes indicate WT amino acids, dark blue boxes indicate variant side chains. The figure illustrates the hypothetical double mutation G39A-I189Q.}
\label{fig:assm_mutn}
\end{figure}

After introducing the mutation, the atoms of the new side chain are adjusted by molecular modeling to be in overlay with the wild type side chain and to fit into the available geometrical space.  Each amino acid of the protein is then stored into a separate PDB file (called "fragment", (1) in Fig. \ref{fig:assm_mutn}). The water molecules and the substrate are stored as separate PDB files as well. By substituting the PDB fragment of the wild type at a given position by the fragment PDB file of a mutated side chain, the PDB structure file of a mutated enzyme can be assembled ((2) in Fig. \ref{fig:assm_mutn}).

\begin{figure}[!ht]
\begin{center}
\begin{minipage}{0.32\linewidth}
\includegraphics[width=0.99\linewidth]{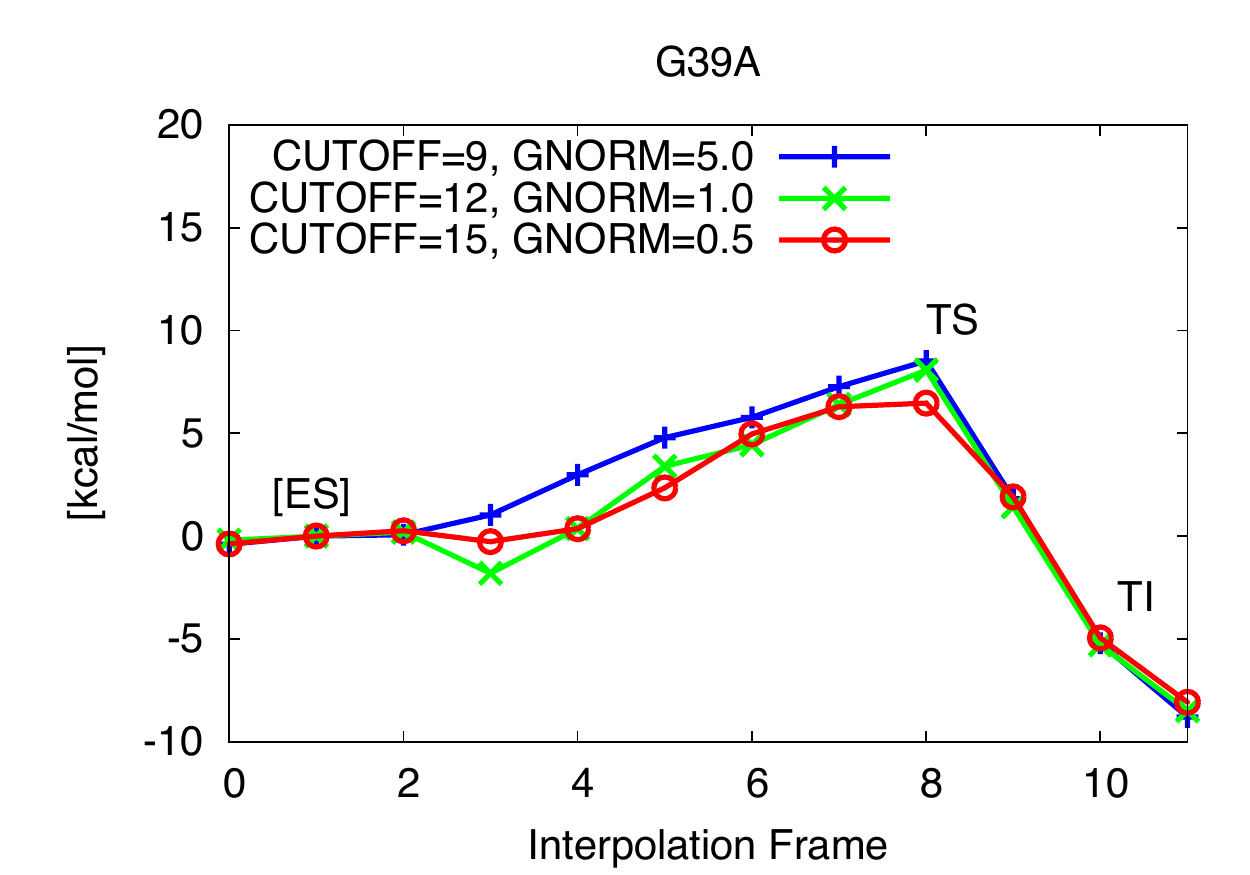}
\end{minipage}
\begin{minipage}{0.32\linewidth}
\includegraphics[width=0.99\linewidth]{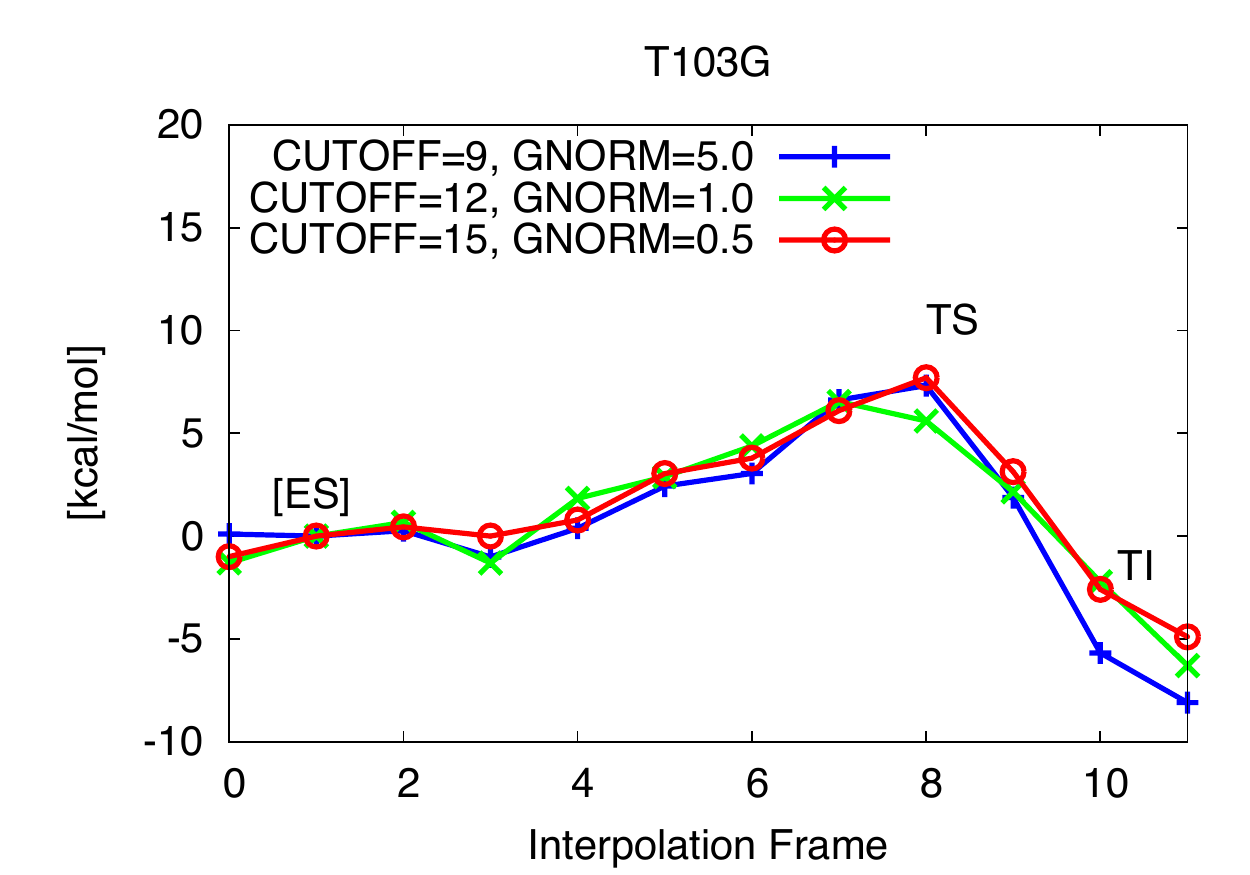}
\end{minipage}
\begin{minipage}{0.32\linewidth}
\includegraphics[width=0.99\linewidth]{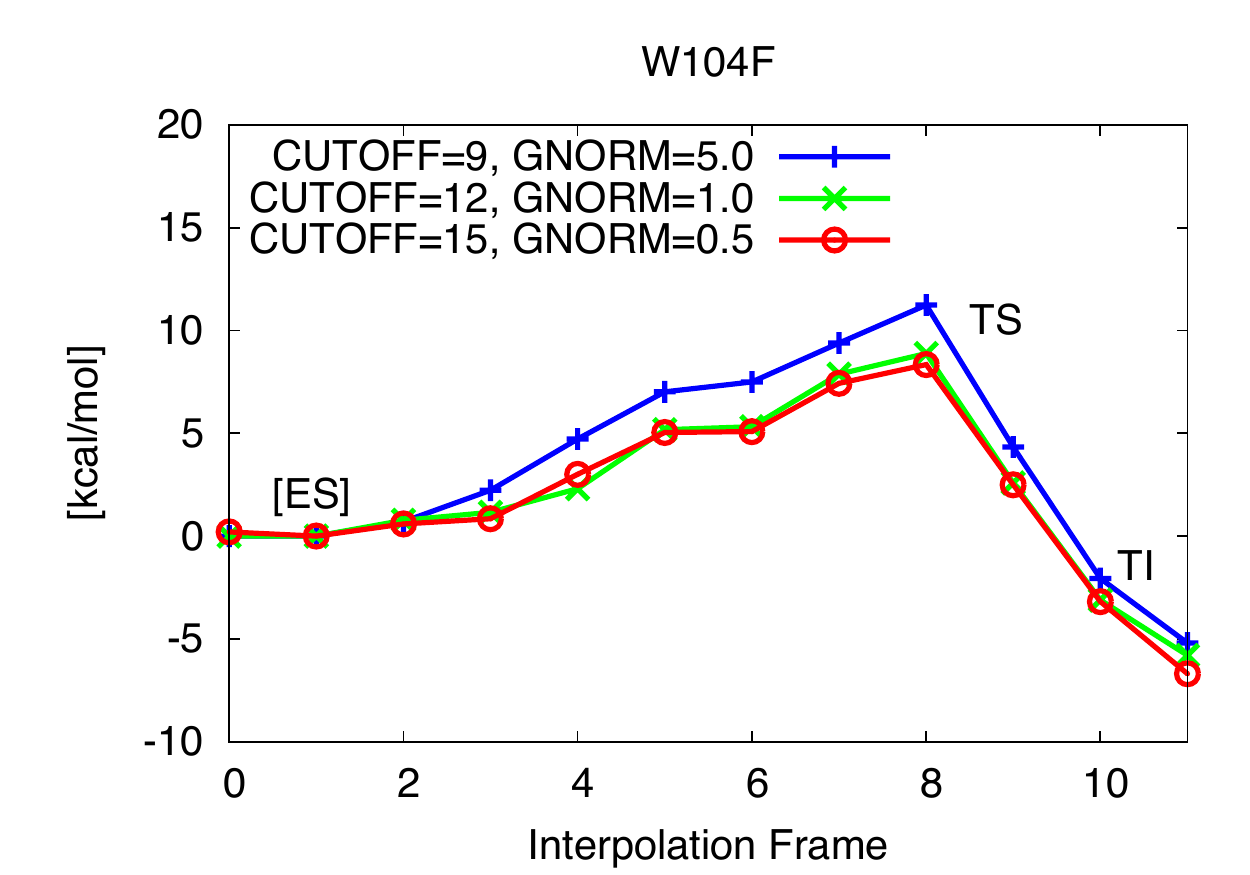}
\end{minipage}
\end{center}
\caption{{\bf Single mutation screening between \textbf{TS} and \textbf{ES} in (\textbf{b})}. { E}lectronic energy difference not corrected for ZPE. The difference is defined by locating the highest point on the PES and subtracting the energy of the lowest point before it. Energy differences from PM6 SPE calculations of M(C15, G0.5) optimized geometries (in kcal/mol): A. G39A: 6.9. B. T103G: 8.7. C. W104F: 8.3. For comparison (see Fig. \ref{fig:barriers}) WT: 6.0.}
\label{fig:single_mutations}
\end{figure}

In the optimization of the interpolation frames of the variants, it was observed that the introduction of a big side chain in the active site can lead to significant rearrangment of side chains on the surface of the molecular model. From this, the bonding topology between the wild type and the mutant can become significantly different and lead to reaction barrier estimates with unconclusive shapes. It was therefore required to fix the atoms of a number of side chains on the surface of the molecular model to remain in the position of the optimized wild type structure. In particular, the side chains of the residues S50, P133, Q156, L277 and P280 are fixed. Other than the constraints on the atoms of the reaction coordinate (which are removed in frames 0 and 11), these constraintes also remain effective in the optimization of the reaction end points. The reaction barrier estimations obtained after carrying out the linear interpolation and the constrained optimization of the variant structures are presented in Fig. \ref{fig:single_mutations}.

The reaction energy profile of the G39A mutant shows a slight decrease in energy at interpolation frame 3. This is explained by the presence of a local minima with lower energy than the initial \textbf{ES} state which becomes available to the system at the third step of the interpolation. However, this decrease in energy is not observed in the optimizations using M(C9, G5.0) or M(C15, G0.5). A similar effect is observed in the profile of the T103G mutant, both for the calculations using M(C9, G5.0) and M(C12, G1.0).

\begin{figure}[!ht]
\begin{center}
\includegraphics[width=4in]{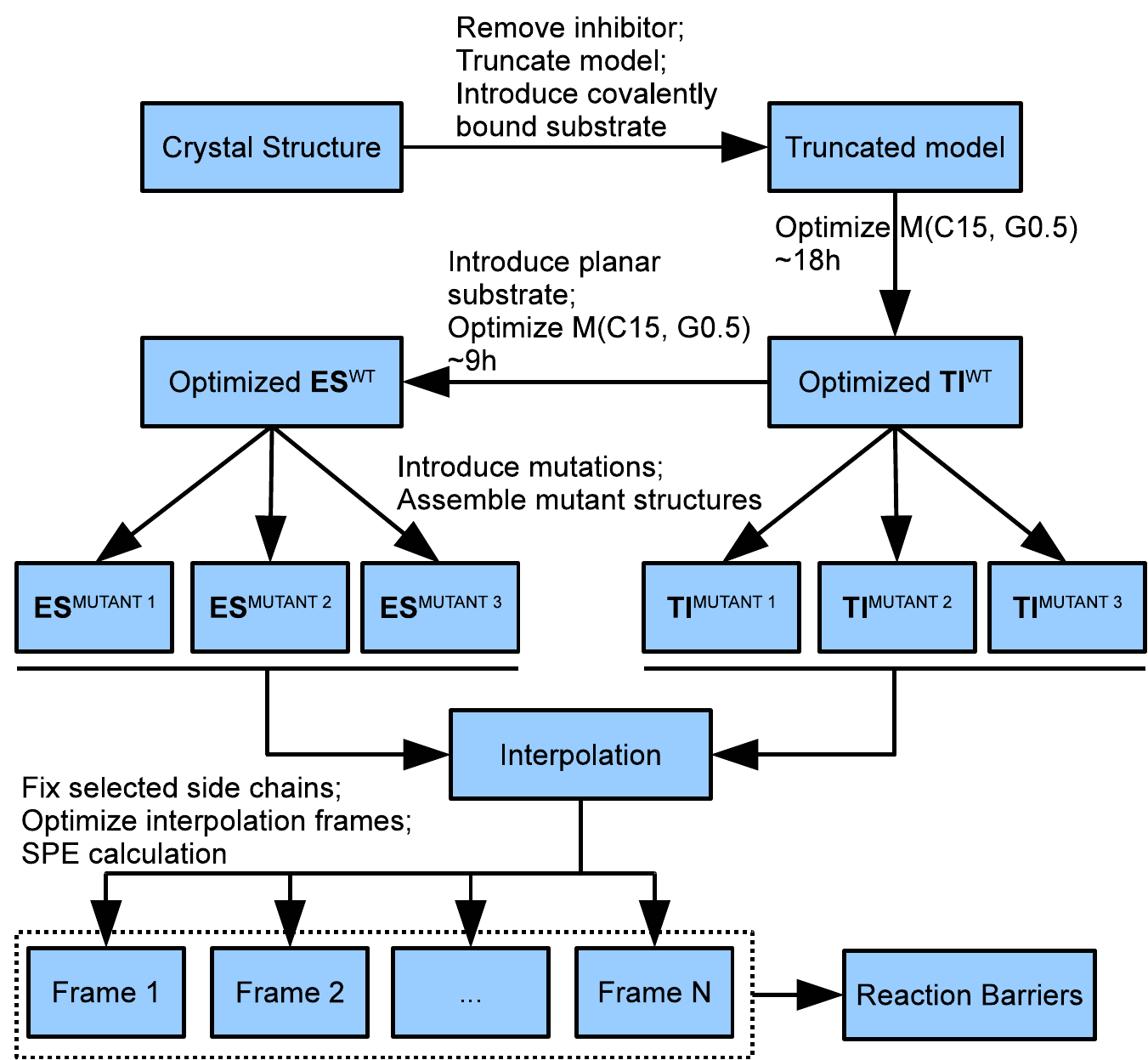}
\end{center}
\caption{{\bf Calculation setup.} { I}nterpolation of WT model (not indicated) is between optimized \textbf{ES} and \textbf{TI} structure.}
\label{fig:calc_path}
\end{figure}

The estimated barrier of the G39A mutant is very close to the WT barrier and the lowest of all three mutants. Based on this, it would be concluded that the G39A mutation is the most likely candidate for showing increased overall activity. The complete approach outlining the various steps included in the presented screening technique is summarized in the overview Fig. \ref{fig:calc_path}.

\subsection*{Time Requirements}\label{sec:sub_cmdt_vrtm}
\begin{figure}[!ht]
\begin{center}
\includegraphics[width=4in]{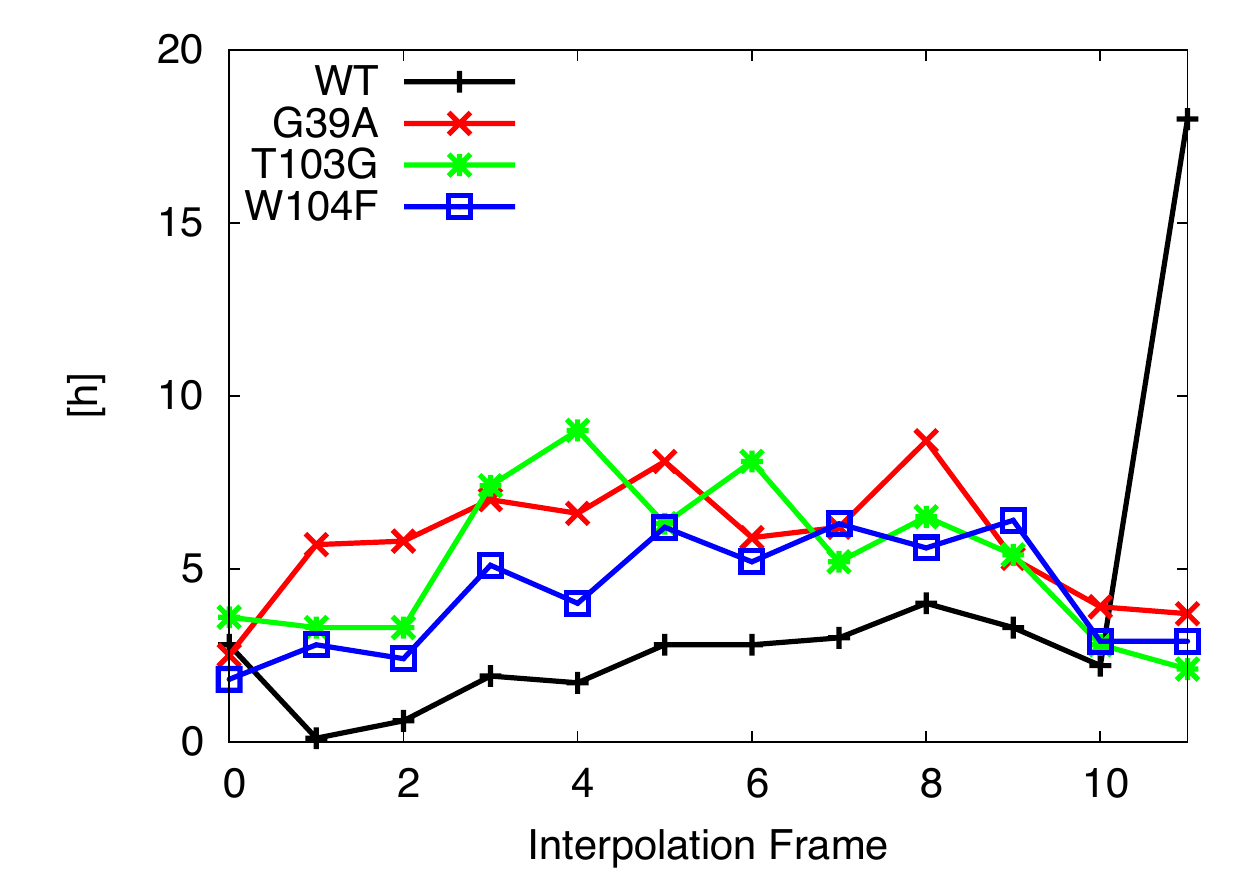}
\end{center}
\caption{{\bf Wall clock time for geometry optimization.} { O}ptimization of \textbf{TI}$^{WT}$ requires more time than mutant \textbf{ES} and \textbf{TI} structures. Average wall clock time per interpolation frame (in h): 4.1, 5.8, 5.2, 4.3 for WT, G39A, T103G and W104F, respectively. All optimizations done using M(C15, G0.5).}
\label{fig:time}
\end{figure}

It was observed that a significant amount of CPU time can be saved by basing the molecular model of the \textbf{ES}$^{WT}$ and the variants on the optimized \textbf{TI} of the wild type. Since the molecular model of the wild type is based on the crystal structure, a major proportion of the structure is already optimized when the mutation is introduced. In Fig. \ref{fig:time}, it is shown how the required wall clock time for the optimization of the wild type and three variants depends on the interpolation frame.

In the figure, a trend towards higher time requirements for the interpolation frames for the non-stationary points is observed. The average time per interpolation frame is highest in the G39A mutation. This appears reasonable considering the fact that a sterically demanding group is being introduced into a restricted environment, which requires considerable rearrangement of the surroundings. The time requirement in all three variants is greatly reduced by basing the molecular model of the variant on the optimized structure of the \textbf{TI} of the wild type. Also, the optimization of frame 1 of the wild type appears to require only very little CPU time (0.1h). This is explained by its high similarity to frame 0, which is completely optimized already.

It is worth noting, that the interpolation frames can be optimized in parallel and thus the CPU time requirement for the evaluation of the energy profile is only determined by the optimization of that interpolation frame with the highest wall clock time.

\subsection*{Conclusions}
A fast computational enzyme activity screening method is presented. The method is designed towards the efficient estimation of the barrier height of an enzymatic reaction of a large number of mutants. Based on the presented approach, the barrier height of a mutant can be computed within 24 hours on roughly 10 processors. In the approach, the PM6 method as implemented in the MOPAC2009 program is used. The approach is tested and applied to the study of the first step of the amide hydrolysis reaction as catalysed by \textit{Candida Antarctica} lipase B (CalB).
In particular we show that

1. PM6 reproduces the \mbox{RHF/3-21G} transition state (\textbf{TS}) structure (Fig. \ref{fig:eval-ts}) and \mbox{B3LYP/6-31G(d)//RHF/3-21G} barrier height (Tab. \ref{tab:barriers}) for a small model system.

2. PM6 combined with the MOZYME method can be used to geometry optimize a structural model containing all residues within 8\AA\ of the active site (Fig. \ref{fig:size_comparison}b) in about 18 hours on a single processor (Tab. \ref{tab:comp_heat}). A gradient convergence criterion of 0.5kcal/(mol\AA) and an NDDO cutoff distance of 15\AA\ are needed for reliable results.

3. The \textbf{TS} search algorithm implemented in MOPAC2009 was found too computationally demanding and not consistently reliable for our purposes. Instead we devised an adiabatic mapping method for estimating the \textbf{TS} structure and barrier height (Fig. \ref{fig:barriers}), where key bond lengths are kept constrained at a series of intermediate values while the rest of the protein structure is optimized using MOZYME. The optimized geometries are then used for conventional (i.e. not MOZYME) PM6 single point energies, because the energy difference between conventional PM6 and MOZYME-PM6 is too large compared with the effect of mutations (Figs. \ref{fig:comp_moz_ort_pm6}, \ref{fig:single_mutations}).

4. The average CPU time needed per point on the energy profile is 4-5 hours on a single processor (Fig. \ref{fig:time}) and each point can be computed independently leading to trivial parallelization (Fig. \ref{fig:calc_path}).

5. Both the preparation of input files for the optimization of
all interpolation frames on the reaction coordinate as well as the generation
of energy profiles are automated to a large degree. In the current setup,
manual effort is required only in the molecular modeling of the mutant side
chain fragment PDB files, Fig. \ref{fig:assm_mutn}, and the molecular modeling
of the substrate in the non-covalently bound reactant state. However, since a
side chain fragment for a given mutant can be used in any number of combination
mutants including this mutation, the required manual effort only scales with
the number of distinguishable point mutations.

The method described here is \textit{in principle} generally applicable to efficiently identify promising mutants for further study for any enzyme-catalyzed reaction for which the structure is known and which does not involve open-shell species (which can not currently be handled with MOZYME).
When applying the method to a new system it is of course important to re-check the validity of using the PM6 method by, for example, comparison to \textit{ab initio} results for small model systems, as was done here.
In addition, the usual caveats associated with \textit{all} computational studies of enzymatic reactivity apply: Identifying a reaction coordinate that uniquely defines the mechanism can be difficult and is ultimately a matter of trial and error.
Mechanisms that involve large structural rearrangements of the enzyme and/or large changes in solvation energy are difficult to model accurately, and the predicted effects of mutations may be less reliable.

As an initial application, the barrier heights of nearly 400 single to four-fold combination mutants in CalB have been estimated and, for 22 mutants, compared to experimentally measured activities with promising results (a preprint of this as yet unpublished study is available at http://arxiv.org/abs/1209.4469).

\section*{Acknowledgments}
The authors acknowledge the \textit{In Silico Rational Engineering of Novel Enzymes} (IRENE) FP7 project for financial support. The authors further acknowledge J. Stewart (Stewart Computational Chemistry) for fruitful discussions.

\bibliography{citations}

\newpage
\section*{Figure Legends}

\clearpage

\section*{Tables}

\begin{table}[!ht]
\caption{
{\bf Energy difference between TS and ES of SE structures.}}
\begin{tabular}{lr}
Energy Differences &  $\Delta$E [kcal/mol]  \\\hline
B3LYP/6-31G(d)//HF/3-21G        & 20.95  \\
B3LYP/6-31G(d)//AM1             & 19.85  \\
B3LYP/6-31G(d)//RM1             & 14.15  \\
B3LYP/6-31G(d)//PM3$^{a}$       &  6.62  \\ 
B3LYP/6-31G(d)//PM6             & 18.20  \\\hline
\end{tabular}
\begin{flushleft}Electronic energy difference between $\textbf{TS}$ and \textbf{ES} in (\textbf{1}), not corrected for ZPE. $a$: PM3 value only for proton abstraction.
\end{flushleft}
\label{tab:barriers}
\end{table} 

\begin{table}[!ht]
\caption{{\bf Comparison of MOPAC configurations and molecular model size.}}
\begin{tabular}{lrrr}
GCC &      PM6//C9 & PM6//C12 & PM6//C15 \\\hline
\multicolumn{4}{l}{(\textbf{a}): 17 Residues}\\
5.0 & -1548.2(0.3) & -1560.1(0.7) & -1561.3(0.5) \\ 
1.0 & -1548.2(0.4) & -1560.1(0.7) & -1561.3(0.5) \\ 
0.5 & -1548.2(0.4) & -1560.1(0.6) & -1561.3(0.7) \\
\multicolumn{4}{l}{(\textbf{b}): 55 Residues}\\
5.0 & -4281.4(2.2) & -4288.6(3.7) & -4286.0(3.1) \\ 
1.0 & -4311.6(7.9) & -4323.5(12.6) & -4317.6(16.1) \\ 
0.5 & -4311.6(7.3) & -4323.5(12.4) & -4318.2(18.0) \\
\multicolumn{4}{l}{(\textbf{c}): 118 Residues}\\
5.0 & -9315.8(18.9) & -9302.5(29.5) & -9296.7(25.0)\\ 
1.0 & -9327.7(37.2) & -9310.3(54.4) & -9306.0(43.6) \\ 
0.5 & -9327.6(37.5) & -9311.8(63.1) & -9316.4(85.8)\\\hline
\end{tabular}
\begin{flushleft}PM6 $\Delta_f$H (in kcal/mol) of \textbf{TI} depending on GCC (in kcal/(mol\AA)) and NDDO cutoff (in \AA), wall clock time for geometry optimization in parenthesis (in h). "PM6//C9": PM6 $\Delta_f$H computed on structure optimized with MOZYME and NDDO cutoff set to 9\AA. Average wall clock time (in h) for SPE calculations in (\textbf{a}), (\textbf{b}) and (\textbf{c}): 0.01, 1.4, 21.0.
\end{flushleft}
\label{tab:comp_heat}
\end{table} 

\section{A Computational Methodology to Screen Activities of Enzyme Variants: Supplementary Material}
Martin Hediger, Luca De Vico, Allan Svendsen, Werner Besenmatter, Jan H. Jensen. Corresponding Author, Email: jhjensen@chem.ku.dk

\subsection{Difference MOZYME/PM6 Energies}
\begin{figure}[htbp]
\centering
\includegraphics[width=0.80\linewidth]{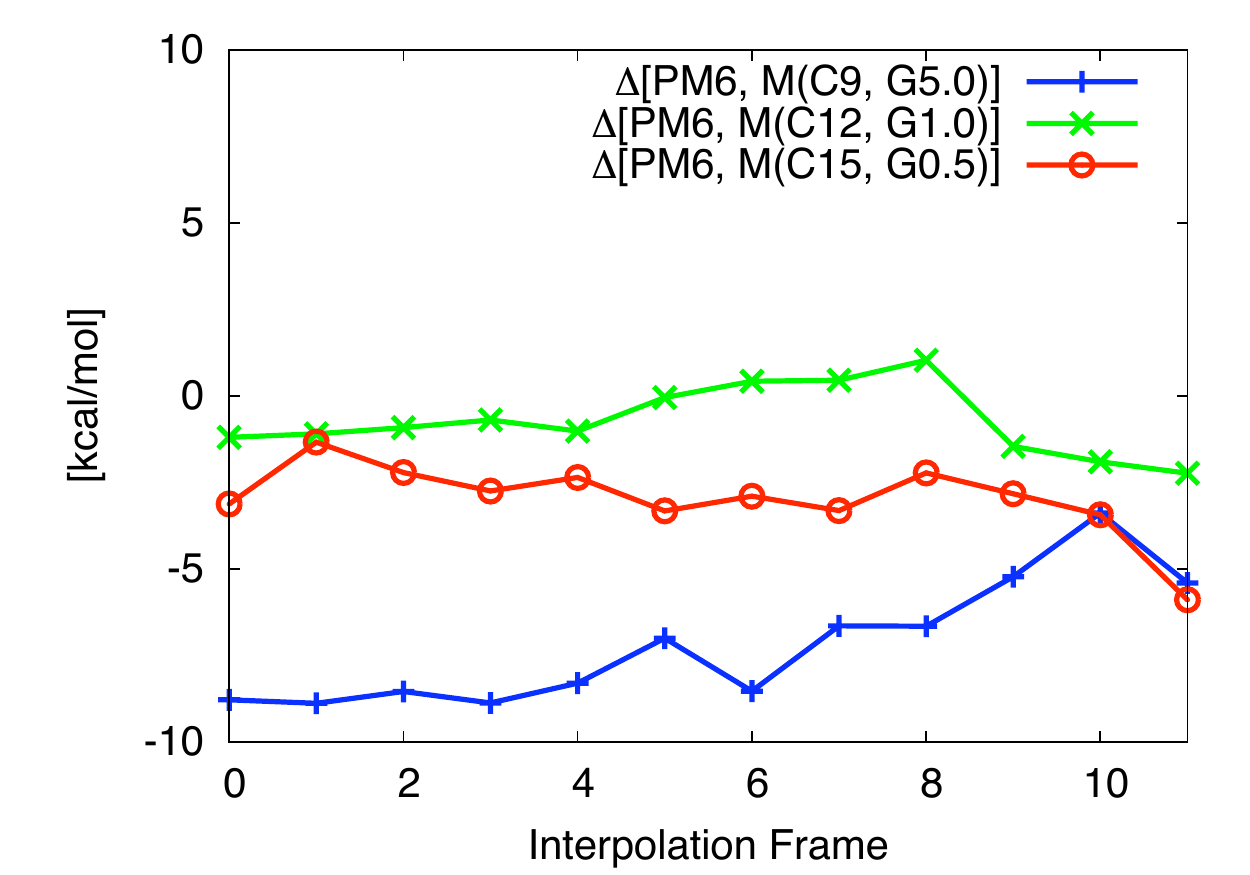}
\caption{Difference $\Delta$H$_f^{PM6}$ - $\Delta$H$_f^{MOZYME}$ between PM6 and MOZYME (not reorthogonalized) energy at each interpolation frame in (\textbf{b}). Average difference (in kcal/mol): -7.19, -0.73, -2.98 for NDDO cutoff 9, 12, and 15\AA, respectively.}
\label{fig:supp_comp_pm6_moz}
\end{figure} 

\subsection{Amino Acids in Models (\textbf{a}), ($\textbf{b}$) and (\textbf{c})}
All amino acids in the indicated ranges included in the respective models.
\\
(\textbf{a}): 38-42, 103-106, 134, 187, 189, 223-225, 278, 281\\ 
(\textbf{b}): 37-50, 102-106, 131-135, 139-141, 156-158, 186-192, 220-226, 277-287\\ 
(\textbf{c}): 35-52, 66-83, 101-114, 128-141, 153-158, 163-167, 172-174, 179-193, 200-204, 223-233, 277-285

\subsection{Transition State Verification}
The animation of the vibration can be found online under:\\
\verb+http://www.youtube.com/watch?v=7ZLaqH2xDy8+

\subsection{Git Repository for Scripts}
The scripts used in the development of this method are available under:\\
\verb+git@github.com:mzhKU/Enzyme-Screening.git+

\end{document}